\documentclass[final]{IEEEtran}

\usepackage{cite}
\usepackage{amsmath}
\usepackage{color}
\usepackage{theorem}
\usepackage{subfigure}
\usepackage[lined,algonl,boxed,ruled,linesnumbered]{algorithm2e}
\usepackage{psfrag}
\usepackage{amsbsy}
\usepackage{amssymb}
\usepackage[mathscr]{eucal}
\usepackage{latexsym}
\usepackage{bm}
\usepackage{array}
\usepackage{dsfont}
\usepackage{mfirstuc}
\DeclareMathAlphabet{\mathpzc}{OT1}{pzc}{m}{it}

\usepackage[acronym,shortcuts]{glossaries}
\usepackage{soul}
\usepackage{tikz}
\usepackage{pgf}
\usepackage{multirow}
\usepackage{comment}
\usepackage{colortbl}

\usepackage{todonotes}
\let\svtodo\todo\renewcommand\todo[1]{\svtodo[inline]{#1}}

\newacronym{BPSK}{BPSK}{binary phase shift keying}
\newacronym{CSI}{CSI}{channel state information}
\newacronym{DEC}{DEC}{decoder}
\newacronym{ENC}{ENC}{encoder}
\newacronym{iid}{iid}{independent identically distributed}
\newacronym{MIMO}{MIMO}{multiple input multiple output}
\newacronym{MRC}{MRC}{maximal ratio combining}
\newacronym{NACK}{NACK}{not acknowledge}
\newacronym{PDF}{PDF}{probability density function}
\newacronym{SNR}{SNR}{signal-to-noise ratio}
\newacronym{CER}{CER}{codeword error rate}
\newacronym{OFDM}{OFDM}{orthogonal frequency division multiplexing}
\newacronym{DF}{DF}{decode and forward}
\newacronym{AF}{AF}{amplify and forward}
\newacronym{PLA}{PLA}{physical layer authentication}
\newacronym{PLS}{PLS}{physical layer security}
\newacronym{CIR}{CIR}{channel impulse response}
\newacronym{AWGN}{AWGN}{additive white Gaussian noise}
\newacronym{OCC}{OCC}{one class classification}
\newacronym{TPR}{TPR}{true positive rate}
\newacronym{TNR}{TNR}{true negative rate}
\newacronym{NN}{NN}{nearest neighbor}
\newacronym{OCNN}{OCNN}{One-class Nearest Neighbor}
\newacronym{GLRT}{GLRT}{generalized likelihood ratio test}
\newacronym{c.d.f.}{c.d.f.}{cumulative density function}
\newacronym{p.d.f.}{p.d.f.}{probability density function}
\newacronym{ML}{ML}{maximum likelihood}
\newacronym{LLR}{LLR}{logarithm of the likelihood ratio}
\newacronym{SVM}{SVM}{support vector machine}
\newacronym{OC-SVM}{OC-SVM}{one-class support vector machine}
\newacronym{RBF}{RBF}{Radial Basis Function}
\newacronym{IoT}{IoT}{Internet of Things}
\newacronym{FA}{FA}{false alarm}
\newacronym{MD}{MD}{missed detection}

\definecolor{ashgrey}{rgb}{0.7, 0.75, 0.71}

\begin{document}

\title{Comparison of Statistical and Machine Learning Techniques for Physical Layer Authentication}

\author{Linda Senigagliesi, Marco Baldi and Ennio Gambi
\thanks{The material in this paper was presented in part at the IEEE Global Communications Conference (Globecom 2019), Waikoloa, HI, USA, Dec. 2019, and at the IEEE International Workshop on Information Forensics and Security (WIFS 2019), Delft, The Netherlands, Dec. 2019.
}
\thanks{L. Senigagliesi, M. Baldi and E. Gambi are with Dipartimento di Ingegneria dell'Informazione, Universit\`a Politecnica delle Marche, 60131 Ancona, Italy (e-mail: l.senigagliesi@staff.univpm.it, m.baldi@univpm.it, e.gambi@univpm.it).}
}

\maketitle

\begin{abstract}
In this paper we consider authentication at the physical layer, in which the authenticator aims at distinguishing a legitimate supplicant from an attacker on the basis of the characteristics of a set of parallel wireless channels, which are affected by time-varying fading. Moreover, the attacker's channel has a spatial correlation with the supplicant's one.
In this setting, we assess and compare the performance achieved by different approaches under different channel conditions.
We first consider the use of two different statistical decision methods, and we prove that using a large number of references (in the form of channel estimates) affected by different levels of time-varying fading is not beneficial from a security point of view.
We then consider classification methods based on machine learning. In order to face the worst case scenario of an authenticator provided with no forged messages during training, we consider one-class classifiers.
When instead the training set includes some forged messages, we resort to more conventional binary classifiers, considering the cases in which such messages are either labelled or not.
For the latter case, we exploit clustering algorithms to label the training set.
The performance of both nearest neighbor (NN) and support vector machine (SVM) classification techniques is evaluated. Through numerical examples, we show that under the same probability of false alarm, one-class classification (OCC) algorithms achieve the lowest probability of missed detection when a small spatial correlation exists between the main channel and the adversary one, while statistical methods are advantageous when the spatial correlation between the two channels is large.
\end{abstract}



\section{Introduction}

Classical authentication protocols rely on cryptographic primitives to allow the supplicant to prove his identity to the authenticator.
Although such approaches are well consolidated and implemented in real world applications, they may exhibit some limitations when adopted in new scenarios, like that of the \ac{IoT}.
In fact, cryptographic approaches are usually characterized by a relatively large complexity, since security relies on the difficulty for a computationally-constrained attacker to break some mathematical trapdoor.

In \ac{IoT} applications, huge numbers of resource-constrained cyber-physical devices are deployed, and they need to be authenticated in order to avoid impersonation and falsification attacks.
In such a context, the use of classic authentication protocols based on cryptographic primitives may result cumbersome, due to computing power and memory constraints of embedded devices.
Opposed to approaches based on cryptographic primitives, \ac{PLS} approaches exhibit some features that make them particularly suitable for \ac{IoT} applications, that is:
\begin{itemize}
    \item they do not require any assumption on the computing power of attackers;
    \item they only rely on the communication channel's characteristics and do not need pre-shared credentials;
    \item they are characterized by low complexity.
\end{itemize}
For this reason, a recent trend in the literature is focused on \ac{PLS} solutions for \ac{IoT} applications \cite{Mukherjee2015, Sun2018, Hindawi2019,Wang2019}. Low-complexity security frameworks for key generation based on \ac{PLS} have also been proposed \cite{Hanzo2019}, together with solutions that consider adversaries having infinite computational capabilities \cite{Wang2016b} over realistic wireless channels \cite{Baldi2017}.

Authentication is one of the first and most important tasks in secure communications.
It aims at recognizing messages coming from a legitimate supplicant while detecting those which may be forged by a malicious attacker. 
At the physical layer, authentication is performed by distinguishing the source of a message based on the unique characteristics of the communication channel \cite{Caparra2017}.

In this paper, we consider and design some \ac{PLA} protocols based on several decision criteria, both based on statistical hypothesis testing and machine learning approaches.
We consider wireless parallel channels affected by time-varying fading to assess the performance of the \ac{PLA} protocols we examine. The legitimate receiver does not know the exact realizations of the channel between the transmitter and itself, but he may know some of its statistical properties, such as the noise variance, or even not have any \ac{CSI}. In the latter case, authentication must hence be performed blindly with respect to the
channel characteristics.
Concerning decision methods, we start from the statistical techniques used in \cite{Baracca2012} for a flat fading wireless channel model, and consider several others decision criteria.
We also start from statistical criteria based on a hypothesis testing approach \cite{Maurer2000b}, and their corresponding optimal attacks. 
Then we consider decision criteria based on machine learning, whose application to \ac{PLA} has started to attract interest in the literature in the last years under different scenarios \cite{Weinand2017, Wang2017}. 
For this purpose, we consider several classification methods based on machine learning, and compare the performance they achieve with that obtainable through classical statistical methods.
In particular, we use \ac{NN} and \ac{SVM} algorithms, examining both their binary and one-class versions. The former have a low computational complexity with a reduced training set, which makes them attractive in resource-constrained applications, while the latter allow avoiding the search of optimal parameters in the initial phase. 
We perform the training phase considering both the case of an authenticator who knows exactly the source of the received setup packets and that of an uninformed authenticator. We therefore also consider clustering algorithms to establish the identity of the message sender during the initial phase.
One of the points raised in [17] is the difficulty in pre-designing a precise authentication model. By exploiting predefined algorithms we can assess the performance of generic channel models, not specifically designed for the examined scheme.
In order to have a shared and reproducible benchmark, we consider the use of off-the-shelf machine learning algorithms. A possible follow-up of our analysis is the design of specific classifiers for the considered application, but this is out of the scope of the present work and is left for future works.

\subsection{Related work}

Blind authentication schemes have been proposed in the last years, as in \cite{Xie2018}, where techniques of blind known-interference cancellation and differential processing are combined to implement authentication, but they rely on the sharing of a secret key between legitimate participants.
Effects of time-varying fading have been studied in literature considering schemes based both on the sharing of a secret key \cite{Xie2018} and key-less approaches \cite{Xiao2008, Tugnait2014}.
To the best of our knowledge, machine learning has been applied to \ac{PLA} only very recently and under different conditions, for instance not considering time-varying fading channels (see \cite{Pan2017}, \cite{Wang2017}). A review of machine-learning-aided intelligent authentication techniques proposed for 5G communications, with the definition of the requirements needed for new generation networks is provided in \cite{Fang2019b}. In \cite{Baldini2018}, $k$-\ac{NN} algorithms are used to identify electronic devices through their radio-frequency emissions, and the presence of an attacker is not explicitly considered. One-class \ac{SVM} and k-means clustering have been successfully applied in a context different from the one we consider, such as in \ac{MIMO} stationary systems \cite{Yoon2019}, or for the detection of eavesdropping attacks in unmanned aerial vehicle (UAV)-aided wireless systems \cite{Hoang2020}.
Besides considering scenarios different from ours, all the mentioned works also consider the use of high level cryptographic schemes during the first phase of the authentication process.
Authors of \cite{Fang2019} propose the application of an adaptive algorithm for \ac{PLA} in a dynamically changing wireless environment, considering a single channel model and a series of physical layer attributes which may include \ac{CSI}.
Very recently, the use of adaptive neural networks to achieve adaptive authentication has been proposed in \cite{Qiu2020}, also including a prototype implementation on a universal software radio peripheral (USRP) platform in presence of multipath fading. 
However, in \cite{Qiu2020} it is assumed that a pre-shared secret key is available during the training phase, while we exploit clustering algorithms to perform authentication without either shared secrets or previous knowledge on the legitimate transmitter. Furthermore, although the methodology proposed in \cite{Qiu2020} can be implemented on USRP platforms, the use of neural networks is not suitable for resource-constrained scenarios, as the one of interest in this work.

\subsection{Contribution}

We start from the analysis developed in \cite{Baracca2012}, which considers only flat fading, and extend it to the time-varying fading case. 
We compare the performance obtained for the flat fading case with that achieved with time-varying fading, showing the effect of fading on authentication.
Then, we consider the use of machine learning techniques to accomplish the same task in the same setting, and compare their performance with that of the statistical techniques used in \cite{Baracca2012}.
We consider a training phase corresponding to the initial state of the channel, 
and the existence of some correlation between the main channel and the adversary's one. 
This models a more realistic training phase with respect to \cite{Fang2019}.
In addition, differently from \cite{Fang2019}, we also consider the use of non-supervised techniques to discriminate authentic from forged messages during the training phase, letting a clustering algorithm to decide about the nature of the received packets. We show that even in such an unfavorable condition it is possible to achieve good security performance using machine learning algorithms, especially when the attacker's channel has a low spatial correlation with the main one.

This work consolidates and extends our previous analyses reported in \cite{Senigagliesi2019, Senigagliesi2019b} by:
\begin{itemize}
    \item Exploiting clustering algorithms to establish the source of a message during the initial training phase, in order to avoid the need for higher layer cryptographic protocols or manual procedures.
    \item Comparing different machine learning techniques based on different classification principles, also considering their binary and one-class versions.
    \item Providing a more extensive comparative assessment between statistical and machine techniques applied to \ac{PLA} under different conditions.
    Introducing a hybrid approach that embeds statistical metrics into machine learning algorithms. In fact, we show that statistical metrics can be used with \ac{NN} algorithms to improve the ability to detect forged messages.
    \item Considering the more general case in which fading also affects the training phase, showing that this leads to a loss in terms of missed detection with both the analyzed approaches.
    \end{itemize}

The paper is organized as follows. In Section \ref{sec:sys_model} we introduce the authentication protocol and the channel model considered, along with the relevant performance metrics. 
In Section \ref{sec:stat_methods} two different decision methods based on statistical techniques are introduced. 
Section \ref{sec:attacker} describes the attacker models, while Section \ref{sec:machine_learning} describes the one-class classification algorithms we apply. 
Numerical evaluations and the relevant results are reported in Section \ref{sec:results}, after which Section \ref{sec:concl} provides some conclusive remarks.

\section{System model and metrics}
\label{sec:sys_model}

The following channel model is considered. A peer (Alice) has to be authenticated by an authenticator (Bob) at the presence of a malicious attacker (Eve), who aims at impersonating Alice by forging her messages. Through \ac{PLA}, Bob should be able to recognize the messages coming from Alice as legitimate and to refuse the ones coming from Eve.

Transmission of a message is performed over a set of $N$ parallel channels, which can model multi-carrier or \ac{OFDM} transmission. 
Each channel is corrupted by \ac{AWGN} and time-varying fading, which represents a standard model used in the literature to assess and compare the performance of different transmission techniques.
The rapidity of channel variations clearly influences the quality of authentication, since it makes more difficult for Bob to recognize channel estimates coming from the same source. 
If we wish to take into account the possible occurrence of interference, we must broaden the study and include network aspects, e.g., concerning the number of nodes, the network topology and the adopted medium access control protocols. Although this may be interesting in principle, it opens the way to many degrees of freedom, and makes benchmarks susceptible to the chosen network parameters and protocols. This is clearly out of the scope of this paper, thus we leave it for future works.

Channel samples representing the \ac{CIR} are collected into a vector $\mathbf{h}$ of complex numbers, whose entries are zero-mean correlated circularly symmetric complex Gaussian variables.
Each channel estimate is then written as
\begin{equation}
\mathbf{h}^{(\textsc{XY})} \sim \mathcal{CN} (\mathbf{0}_{N\times 1}, \mathbf{R}^{(\textsc{XY})}),
\label{eq:hvec}
\end{equation}
where $X$ and $Y$ represent a general couple of transmitter and receiver, respectively, and $\mathcal{CN} (\mathbf{0}, \mathbf{R})$ denotes the distribution of circularly symmetric complex Gaussian random vectors (with zero mean) having covariance matrix $\mathbf{R}$.

The authentication procedure we consider is based only on channel estimates and comprises two phases, that are summarized next.

\paragraph{Phase I (training)} 

Bob observes one or more packets with known content transmitted from Alice during a fixed interval of time. The source of the messages can be guaranteed either by exploiting higher layer protocols or through physical measures (e.g., by manually executing the setup phase). By exploiting these setup packets, Bob obtains a set of $M$ time-correlated channel estimates which, depending on the duration of the time interval, may be subject to time-varying fading. The $m$-th estimate can be written as
\begin{equation}
    \hat{\mathbf{h}}^{(\textsc{AB})}(m) = \boldsymbol{\alpha}^{\textsc{(I)}}(m) \mathbf{h}^{(\textsc{AB})} + \sqrt{1-\boldsymbol{\alpha}^{\textsc{(I)}2}(m)}\mathbf{w}(m) + \mathbf{w}^{\textsc{(I)}}(m),
    \label{eq:BobEstimate}
\end{equation}
where $\mathbf{w}^{\textsc{(I)}} \sim \mathcal{CN} (\mathbf{0}_{N\times 1}, \sigma_{\textsc{I}}^2\mathbf{I}_N)$ is a noise vector.
The contribution of time-varying fading is represented by a $[1 \times N]$ vector $\boldsymbol{\alpha}$, whose elements correspond to real numbers taking values in $[0,1]$, and by a random variable $\mathbf{w}$ generated according to a Rayleigh distribution with unitary variance. 
We work under the hypothesis of slowly time-varying fading channels, meaning that the fading coefficient is assumed constant during transmission of each packet \cite{Wiesel2006}.

If during Phase I Bob collects $M > 1$ reference estimates, he can average over them in order to reduce the noise level and to obtain the average value of the fading parameter $\boldsymbol{\alpha}^{\textsc{(I)}}$, which may vary from one estimate to another. Hence he obtains the average estimate  
\begin{equation}
    \bar{\mathbf{h}}^{(\textsc{AB})} = \frac{1}{M} \sum_{m=1}^M \hat{\mathbf{h}}^{(\textsc{AB})}(m) =  \boldsymbol{\bar{\alpha}}^{\textsc{(I)}} \mathbf{h}^{(\textsc{AB})} + \sqrt{1-\boldsymbol{\bar{\alpha}}^{\textsc{(I)}2}}\mathbf{\bar{w}} + \mathbf{\bar{w}}^{\textsc{(I)}},
    \label{eq:BobEstimate_ave}
\end{equation}
where $\boldsymbol{\bar{\alpha}}^{\textsc{(I)}}$, $\mathbf{\bar{w}}$ and $\mathbf{\bar{w}}^{\textsc{(I)}}$ represent the average value of the time-varying fading and noise vectors, respectively.
$\bar{\mathbf{h}}^{(\textsc{AB})}$ can be used as a reference to classify new estimates coming from unknown sources.

\paragraph{Phase II (classification)}

After the training phase has been completed, Bob receives further packets without assurance that they come from Alice.
Bob estimates the channel through which these new packets arrive, and exploits such an estimate to decide their source. 
This is done by comparing any of these estimates with the reference one obtained during the training phase.
As done in \cite{Baracca2012}, we suppose that in this phase Eve can forge packets on which Bob's estimate is forced to be equal to a vector $\mathbf{g}$ (plus noise).

In order for Bob to decide if a packet comes from Alice or from Eve during the classification phase, we consider and compare two approaches:
\begin{itemize}
    \item classical statistical methods based on hypothesis testing;
    \item modern techniques based on machine learning.
\end{itemize}

When Bob resorts to statistical methods, it is convenient for him to use the average estimate in \eqref{eq:BobEstimate_ave} as a benchmark, while when machine learning-based decision criteria are used he can exploit the whole training set deriving from the $M$ channel estimates.
The two approaches are described next.

\subsection{Performance metrics}

The performances achievable through the considered approaches are  evaluated by measuring their probability of \ac{FA} and of \ac{MD}.
By \textit{false alarm} we mean the event that occurs when Bob rejects a message coming from Alice, while there is a \textit{missed detection} when he accepts a message forged by Eve.
If we take a measure to reduce the probability of \ac{FA}, normally this increases the probability of \ac{MD}. Therefore, a trade-off between these two effects has to be found.
It is important to observe that while time-varying fading negatively affects the correct authentication of legitimate signals, it plays a positive role from the attacker's point of view.
In fact, as will be shown in the following, it directly influences the probability of \ac{FA}, forcing Bob to accept a larger range of inputs, and thus increasing the chances for Eve that one of her forged messages is accepted as authentic.

As regards the statistical methods, analytical formulations of the two mentioned probabilities can be found and depend on the performed test. 
Probabilities of \ac{FA} and \ac{MD} resulting from the application of machine learning algorithms can instead be evaluated by exploiting the so-called \textit{Confusion Matrix} \cite{Stehman1997}. 
The confusion matrix provides a comprehensive overview of classification results. Columns represent the predicted values, while rows represent the actual values.
It contains 4 kinds of entries, the true positives, the false positives, the true negatives and the false negatives. The total number of samples in the test data corresponds to the sum of these entries.

False positives (FP) represent samples coming from Eve incorrectly classified as positives, i.e. considered as authentic by Bob. This event corresponds to a \ac{MD}. A large number of FPs can be due to a noisy training set and to the impact of fading in Phase I.
If we define the number of negative samples classified as negative (or true negatives) as TN, the probability of \ac{MD} of a classifier can be written as
\begin{equation}
    P_{\textsc{MD}} = \frac{\textsc{FP}}{\textsc{FP}+\textsc{TN}} = 1 - \textsc{TNR},
    \label{eq:ex_TNR}
\end{equation}
where $\textsc{TNR} = \frac{\textsc{TN}}{\textsc{TN} + \textsc{FP}}$ represents the true negative rate.

False negatives (FN) are instead messages coming from Alice refused by Bob (positive samples classified as negative). This event corresponds to a \ac{FA}. If we denote as $\textsc{TP}$ the number of true positives (positive samples classified as positive), the probability of \ac{FA} corresponds to 
\begin{equation}
P_{\textsc{FA}} = \frac{\textsc{FN}}{\textsc{TP}+\textsc{FN}} = 1-\textsc{TPR},
\label{eq:ex_TPR}
\end{equation}
where $\textsc{TPR} = \frac{\textsc{TP}}{\textsc{TP} + \textsc{FN}}$ represents the true negative rate.

Another common metric used to evaluate the performance of a classification algorithm is the accuracy. It is defined as the ratio between the number of correct predictions and the total number of instances classified or, in formulas
\begin{equation}
    \textsc{Acc} = \frac{\textsc{TP} + \textsc{TN}}{\textsc{TP} + \textsc{FP} + \textsc{TN} + \textsc{FN}}.
    \label{eq:acc}
\end{equation}

Accuracy and other widespread performance metrics of supervised classification, however, are not always suitable for the \acs{OCC} scenario, since negative data have a highly skewed distribution with regard to the target data, and could lead to misleading values.
To overcome this issue, we resort to the \textit{Geometric Mean} ($g_{mean}$) of accuracy (introduced in \cite{Kubat97}), measured separately on each class; by combining the \ac{TPR} and the \ac{TNR}, it is defined as
\begin{equation} \label{eq:gmean}
    g_{mean} = \sqrt{\textsc{TPR} \cdot \textsc{TNR}} = \sqrt{(1-P_{\textsc{FA}})(1-P_{\textsc{MD}})}.
\end{equation}

An important property of $g_{mean}$ is that it is independent of the distribution of positive and negative samples in the test data. 
This allows assessing the classifier performance not only on the basis of the predominant class (as happens for the accuracy), but on both classes.

\section{Statistical methods}
\label{sec:stat_methods}

According to classical statistical decision methods, Bob resorts to hypothesis testing \cite{Kay1993} to decide whether the transmission was performed by Alice or not. Hypothesis testing can be applied only when the authenticator has no information about the statistics of the attacker's channel. The considered channel model is in fact focused on determining if the source of the received message is the authentic one or not, and not on recognizing who is the transmitter. The authenticator, Bob, bases the authentication only on the knowledge of Alice's channel statistics, and the event of Eve sending a message is therefore noticed only when the statistics of the new estimates differ from \eqref{eq:BobEstimate}. Denoting by $\hat{\mathbf{h}}$ the channel estimated by Bob, the two hypotheses hence are

\begin{itemize}
\item $\mathcal{H}_0$: the message is coming from Alice.
The new measured channel estimate is subject to time-varying fading and its correlation with the reference estimates collected during Phase I depends on how severely the fading affects the channel. 
The hypothesis $\mathcal{H}_0$ at time $t$ can therefore be written as
\begin{equation}
    \hat{\mathbf{h}}(t) = \boldsymbol{\alpha}^{\textsc{(II)}}(t)\mathbf{h}^{(\textsc{AB})} + \sqrt{1- \boldsymbol{\alpha}^{\textsc{(II)}2}(t)}\mathbf{w}_F + \mathbf{w}^{(\textsc{II})}(t) ,
    \label{eq:H0_f}
\end{equation}
where the noise vector is written as $\mathbf{w}^{(\textsc{II})} \sim \mathcal{CN} (\mathbf{0}_{N\times 1}, \sigma_{\textsc{II}}^2\mathbf{I}_N)$. In order to distinguish between the impact of time-varying fading in Phase I and Phase II, in this second case it is represented by means of the vector $\boldsymbol{\alpha}^{\textsc{(II)}}$ and of a random variable $\mathbf{w}_F$.

\item $\mathcal{H}_1$: the message is coming from Eve, and
\begin{equation}
    \mathbf{\hat{h}}(t) = \mathbf{g} + \mathbf{w}^{(\textsc{II})}(t).
    \label{eq:H1_f}
\end{equation}
Since information about Eve's channel is not available, the hypothesis $\mathcal{H}_1$ does not have its own statistical model, but it represents the complement of the null hypothesis. We in fact suppose that Eve can forge messages through which she can modify the channel estimation obtained by Bob for the Eve-Bob channel into any possible vector $\mathbf{g}$.
\end{itemize}

When Bob does not know the statistical distribution of the attacker's channel, the presence of multiple attackers always boils down to the case of a single attacker. The authenticator in fact must refuse all new messages that differ from Alice's channel estimate known to him, regardless of which attacker has forged the message.
The general case with more than one legitimate transmitter is not considered here, but it can be derived from the model presented. We omit it for brevity.

Two different statistical criteria for Bob to decide between the two hypotheses $\mathcal{H}_0$ and $\mathcal{H}_1$ are considered next.
The performance achievable through these tests is then evaluated by measuring their probability of \ac{FA} and of \ac{MD}.

\subsection{Logarithm of likelihood ratio test}
\label{sec:llr}

Let us first consider the \ac{GLRT} \cite{Kay1993}, as done in \cite{Baracca2012}, where flat fading channels were considered and $\mathbf{g}$ is replaced by its \ac{ML} estimate. We take into account the more general case in which channel variations occur during the authentication.

The \ac{LLR} of a channel estimate $\hat{\mathbf{h}}$ over its $N$ components can be written as \cite{Baracca2012}:
\begin{equation}
    \Psi \varpropto 2 \sum_{n=1}^{N} {\frac{1}{\sigma_n^2}\left|\hat{h}_n - \bar{h}_n^{(\textsc{AB})} \right|^2},
    \label{eq:psi}
\end{equation}
where $\sigma_n^2$ represents the per-dimension variance, evaluated as $\sigma_n^2 = \sigma_{\textsc{I}}^2 + \sigma_{\textsc{II}}^2 + (1-\bar{\alpha}_n^{\textsc{(I)}2}) + (1- \alpha_n^{\textsc{(II)}2})$. 

By substituting \eqref{eq:H0_f} in \eqref{eq:psi}, we obtain that under the hypothesis $\mathcal{H}_0$, $\Psi$ is a non-central chi-square random variable as in \cite{Baracca2012}, with non centrality parameter 
\begin{equation}
    \mu = \sum_{n=1}^N \frac{2}{\sigma_n^2}\left|(\alpha^{\textsc{(II)}}_n-\bar{\alpha}^{\textsc{(I)}}_n)h_n^{(\textsc{AB})}\right|^2.
    \label{eq:mu}
\end{equation}
We note that $\mu$ is strictly dependent on $\boldsymbol{\bar{\alpha}^{\textsc{(I)}}}$, and becomes zero in the limit case of $\bar{\alpha}^{\textsc{(I)}}_n = \alpha^{\textsc{(II)}}_n = 1$ on each channel (absence of fading during both phases), which boils down to the case considered in \cite{Baracca2012}.

The \ac{GLRT} consists in comparing the LLR with a threshold $\theta > 0$, i.e.
\begin{equation}
 \begin{cases}  \Psi \leq \theta : & \mbox{decide for }\mathcal{H}_0, \\ 
\Psi > \theta : & \mbox{decide for }\mathcal{H}_1.
\end{cases}
\label{eq:threshold}
\end{equation}

We can evaluate the probability of \ac{FA} $P_{\textsc{FA}}$, i.e., the probability that Bob refuses a message coming from Alice, as
\begin{equation}
    P_{\textsc{FA}} = P[\Psi>\theta|\mathcal{H}_0] = 1 - F_{\chi^2,\mu}(\theta),
    \label{eq:P_FA}
\end{equation}
where $F_{\chi^2,\mu}(\cdot)$ denotes the \ac{c.d.f.} of a chi-square random variable with $2N$ degrees of
freedom and noncentrality parameter $\mu$.

By substituting the hypothesis $\mathcal{H}_1$ in \eqref{eq:psi}, we observe that $\Psi$ is again a non-central chi-square random variable, but the noncentrality parameter in this case is given by 
\begin{equation}
    \beta = \sum_{n=1}^N \frac{2}{\sigma_n^2}\left|g_n-\bar{\alpha}^{\textsc{(I)}}_n h_n^{(\textsc{AB})}\right|^2.
    \label{eq:beta}
\end{equation}
We can then calculate the probability of \ac{MD} as
\begin{equation}
    P_{\textsc{MD}} = P[\Psi \leq \theta|\mathcal{H}_1] =  F_{\chi^2,\beta}(\theta).
    \label{eq:P_MD}
\end{equation}
By imposing a target $P_{\textsc{FA}}$, the threshold is set as
\begin{equation}
    \theta = F_{\chi^2,\mu}^{-1}(1-P_{\textsc{FA}}).
    \label{eq:thr}
\end{equation}

In the case we consider, in which fading can also affect the training phase, the authentication performance is always worse than in the case without fading during the training phase. In fact, according to the properties of the non-central chi-squared distribution, the presence of fading (measured by means of the parameter $\alpha^{\textsc{(I)}}$) leads to an increase of both the non-centrality parameters $\mu$ and $\beta$ with respect to their corresponding values in absence of time-varying fading. As a consequence, given the same $P_{\textsc{FA}}$, also the value of the threshold $\theta$ increases, with an inevitable worsening of the probability of MD. This is also proved by the numerical results shown in Sec. \ref{sec:results}B.

\subsection{Ideal performance \label{subsec:bound}}

In order to assess the authentication performance in an ideal setting, let us suppose that in Phase 1 Bob is able to collect messages that surely come from Eve, and thus are in the form
\begin{equation}
    \mathbf{\hat{h}}^{\textsc{(E)}} = \mathbf{g} + \mathbf{w}^{\textsc{(I)}}.
\end{equation} 
In this case we do not need to substitute Eve's forged vector $\mathbf{g}$ with its \ac{ML} estimate, since Bob knows it exactly except for the presence of \ac{AWGN} noise. 
In this case, we can no longer resort to hypothesis testing, since we now have two different hypotheses about the source of the message. However, with a slight abuse of notation, we still use the labels $\mathcal{H}_0$ and $\mathcal{H}_1$ for identifying the two events corresponding to Alice's and Eve's transmissions, respectively.
Thus the \ac{LLR} on the new estimated channel $\mathbf{\hat{h}}$ becomes
\begin{equation}
    \bar{\Psi} = \ln \frac{f_{\mathbf{\hat{h}}|\mathcal{H}_1}(\mathbf{\hat{h}})}{f_{\mathbf{\hat{h}}|\mathcal{H}_0}(\mathbf{\hat{h}})},
\label{eq:glrt_eve}
\end{equation}
where $f_{(\mathbf{\hat{h}}|\cdot)}(\mathbf{\hat{h}})$ represents the \ac{p.d.f.} of $\mathbf{\hat{h}}$ under hypothesis $\mathcal{H}_1$ at the numerator and under hypothesis $\mathcal{H}_0$ at the denominator.

Considering that under hypothesis $\mathcal{H}_0$ $\mathbf{\hat{h}}$ is Gaussian distributed around $\mathbf{\hat{h}}^{\textsc{(AB)}}$ with per-dimension variance $\sigma^2$, while under hypothesis $\mathcal{H}_1$ is Gaussian distributed around $\mathbf{\hat{h}}^{\textsc{(E)}}$ with per-dimension variance $\sigma_{\textsc{E}}^2$ (the value of the variance depends on how Eve forges $\mathbf{g}$, which will be described in Section \ref{sec:attacker}), eq. \eqref{eq:glrt_eve} on the $n$-th sub-carrier can be written as
\begin{equation}
    \bar{\Psi} = N\ln{\frac{\sigma}{\sigma_{\textsc{E}}}} + \frac{1}{2\sigma^2}\sum_{n=1}^N{|\hat{h}_n - \bar{h}_n^{\textsc{(AB})}|^2} - \frac{1}{{2\sigma_{\textsc{E}}^2}}\sum_{n=1}^N{|\hat{h}_n - \hat{h}_n^{\textsc{(E})}|^2}.
\end{equation}

We now exploit the following decision criterion:
\begin{equation}
 \begin{cases}  \bar{\Psi} \leq \bar{\theta} : & \mbox{accept}, \\ 
\bar{\Psi} > \bar{\theta} : & \mbox{refuse},
\end{cases}
\label{eq:threshold_bound}
\end{equation}
where $\bar{\theta}$ represents the minimum value of the threshold needed to achieve a given probability of false alarm. 

Differently from \eqref{eq:P_FA} and \eqref{eq:P_MD}, in this case no closed form expression is available for $P_{\textsc{FA}}$ and $P_{\textsc{MD}}$ and they are estimated through Monte Carlo simulations.
The achievable probabilities of \ac{FA} and {MD} represent a lower bound on the test performance obtainable when applying the \ac{LLR} test and will be used as a benchmark in some examples presented in Section \ref{sec:results}.

\subsection{Combined test\label{sec:CombinedTest}}

As it results from our numerical simulations, reported in Section \ref{sec:results}, the \ac{LLR} test alone however is not sufficient to guarantee a correct authentication (and a small probability of \ac{MD}) for values of $\boldsymbol{\alpha}^{\textsc{(II)}}$ that are not next to $1$.
In order to address this issue and improve performance, we also consider a slightly modified decision strategy for Bob, based on a double verification.
For this purpose, let us consider that Bob still exploits the \ac{LLR} test, but followed by a modulus comparison, to decide whether a message is coming from Alice or Eve. 
The additional test based on the modulus is performed by comparing the modulus of the reference estimate $\bar{\mathbf{h}}^{(\textsc{AB})}$ and the current estimate $\hat{\mathbf{h}}$. Thus we define
\begin{equation}
    \Gamma = \sum_{n=1}^N \left(\left|\bar{h}_n^{(\textsc{AB})}\right|- \left|\hat{h}_n\right|\right).
    \label{eq:phi}
\end{equation}

Using such a simple modulus comparison alone results in a poor performance.
However, we consider that Bob uses both the criterion based on the \ac{LLR} and that based on the modulus: only if both these conditions are met, then Bob accepts the message as authentic.
The verification condition can be therefore written as
\begin{equation}
 \begin{cases} \Phi \leq \theta, -\epsilon \leq \Gamma \leq \epsilon : & \mbox{decide for }\mathcal{H}_0, \\
 \mbox{else }: & \mbox{decide for }\mathcal{H}_1,
\end{cases}
\label{eq:sec_ver}
\end{equation}
where $\epsilon$ is a sufficiently small threshold. In the ideal case of absence of noise and fading during the training phase, $\epsilon$ should be zero. 
Since we are considering a realistic scenario affected by both disturbances, we must allow $\epsilon$ to be greater than zero in order to allow Bob to accept messages coming from Alice.

The probability of false alarm can be therefore defined as the probability that at least one of the two conditions is not verified when the sender is Alice (hypothesis $\mathcal{H}_0$), i.e.
\begin{equation}
    P_{\textsc{FA}} = 1 - P\left[\Psi \leq \theta, -\epsilon \leq \Gamma \leq \epsilon | \mathcal{H}_0 \right],
    \label{eq:Pfa2}
\end{equation}
while the probability of missed detection can be written as
\begin{equation}
    P_{\textsc{MD}} = P\left[\Psi \leq \theta, -\epsilon \leq \Gamma \leq \epsilon | \mathcal{H}_1 \right].
\label{eq:Pmd2}
\end{equation}

Being $\Gamma$ computed as the difference of the modulus of two complex normal random variables, which follow an Hoyt distribution \cite{Hoyt1947}, its \ac{c.d.f.} can be evaluated according to \cite[eq. (8)]{Paris2009}. However, a closed form expression for the joint probability distribution of $\Psi$ and $\Gamma$ is not known, thus for its estimation we resort to Monte Carlo simulations.

\textit{Thresholds optimization:}
In order to find the optimal values $\theta^*$ and $\epsilon^*$ of the thresholds $\theta$ and $\epsilon$, we look for their joint values which minimize the probability of MD, ieeei.e.
\begin{equation}
  (\theta^*,\epsilon^*) = \arg\min_{\theta, \epsilon} P\left[\Psi \leq \theta, -\epsilon \leq \Gamma \leq \epsilon | \mathcal{H}_1 \right],
\end{equation}
under the constraint of a fixed $P_{\textsc{FA}}$.

We exploit a two-steps optimization procedure, which computes the couples $(\theta, \epsilon)$ that satisfy the constraint imposed on $P_{\textsc{FA}}$ and then select the one that gives the minimum $P_{\textsc{MD}}$.

\section{Attacker model}
\label{sec:attacker}

We assume that Eve aims at performing a tailored attack based on Bob's decision strategy. According to well-known Kerckhoffs's principle, we build our system model not relying on the principle of ``security through obscurity'', meaning that the attacker has all the information available. In this sense, the assumption that Eve knows exactly which test is used by Bob is made to model the worst case scenario and consequently adapt the system parameters to prevent possible authentication errors (for example by adjusting the transmission power and the \ac{SNR} on the different channels).

We suppose that she has partial \ac{CSI}, that is, she knows the statistics of all transmission channels, but not the exact channel realizations.
We assume that Eve cannot know the channel estimates made by Bob exactly (this would be possible only if they were in the same position), but she can obtain a slightly modified version of them by eavesdropping the communication channels used by Alice and Bob.
We also suppose that Eve can observe transmissions from Alice to Bob and vice versa, thus estimating $\mathbf{h}^{(\textsc{AE})}$ and $\mathbf{h}^{(\textsc{EB})}$. 
We denote the $m$-th channel estimates obtained by Eve as
\begin{equation}
   \hat{\mathbf{h}}^{(\textsc{AE})}(m) = \rho_{\textsc{AE}}\mathbf{h}^{(\textsc{AB})} + \sqrt{1-\rho_{\textsc{AE}}^2}\mathbf{r}(m) + \mathbf{w}^{(\textsc{AE})}(m), 
   \label{eq:Eve_est1}
\end{equation}
\begin{equation}
    \hat{\mathbf{h}}^{(\textsc{EB})}(m) = \rho_{\textsc{EB}}\mathbf{h}^{(\textsc{AB})} + \sqrt{1-\rho_{\textsc{EB}}^2}\mathbf{r}(m) + \mathbf{w}^{(\textsc{EB})}(m), 
\label{eq:Eve_est2}
\end{equation}
where $\mathbf{w}^{(\textsc{AE})} \sim \mathcal{CN} (\mathbf{0}_{N\times 1}, \sigma_{\textsc{AE}}^2\mathbf{\textsc{I}}_N)$ and $\mathbf{w}^{(\textsc{EB})} \sim \mathcal{CN} (\mathbf{0}_{N\times 1}, \sigma_{\textsc{EB}}^2\mathbf{\textsc{I}}_N)$ represent the noise vectors and $\mathbf{r}$ is a complex normal random vector with unitary variance.
The coefficient $\rho_{\textsc{XY}} \in \left[0;1\right]$ denotes the spatial correlation between two channels linking a generic node $Z$ to two distinct nodes $X$ and $Y$.

Opposed to the correlation of several realizations of the same channel in time, we denote this correlation as spatial correlation of two different channels at some fixed time.
The correlation of the estimates
$\hat{\mathbf{h}}^{(\textsc{AE})}(m)$ and $\hat{\mathbf{h}}^{(\textsc{EB})}(m)$ obtained through \eqref{eq:Eve_est1} and \eqref{eq:Eve_est2}, respectively, with the main channel coefficient $\mathbf{h}^{(\textsc{AB})}$ strictly depends on Eve's position with respect to Bob's one, and is represented by means of the parameters $\rho_{\textsc{AE}}$ and $\rho_{\textsc{EB}}$. If $\rho_{\textsc{AE}} = 1$, for example, Eve and Bob are in the same position and the channels estimated by them with respect to messages transmitted by Alice are identical. On the contrary, when $\rho_{\textsc{AE}}$ tends to zero the channels Alice-Bob and Alice-Eve are completely uncorrelated.

Eve's attack is supposed to be based on the average of the estimates collected in Phase 1, since Bob relies on this for the subsequent authentication phase. As already done by Bob, if she can retrieve $M > 1$ observations of $\mathbf{h}^{(\textsc{AE})}$ and $\mathbf{h}^{(\textsc{EB})}$, she can try to refine her attack by averaging over them in order to reduce the noise level.
This is opposed to a differential authentication approach \cite{Tomasin2018}, in which Bob progressively updates its reference estimate.
The latter, however, is suitable in the case of correlated fading over time, which is not the case we consider.
As a consequence, the time coefficient vector $\boldsymbol{\alpha^{\textsc{(II)}}}(t)$ in Phase 2 does not affect her forged vector $\mathbf{g}$.

\subsection{Attack to the LLR test}

When Bob uses the \ac{LLR} test, Eve's best attack strategy is represented by the \ac{ML} estimate of $\hat{\mathbf{h}}^{(\textsc{AB})}$ based on her observations $\hat{\mathbf{h}}^{(\textsc{AE})}$ and $\hat{\mathbf{h}}^{(\textsc{EB})}$.
According to \cite[eq. (45)]{Baracca2012}, the components of the forged vector $\mathbf{g}$ can be written as
\begin{equation}
    g_n = \hat{h}_n^{(\textsc{EB})}C_n + \hat{h}_n^{(\textsc{AE})}D_n ,
\label{eq:att1}
\end{equation}
with $C_n$ and $D_n$ defined as
\begin{subequations}
\begin{equation}
C_n = \frac{\rho_{\textsc{EB}}\omega_n^{(\textsc{EB})} - \rho_{\textsc{AB}}\rho_{\textsc{AE}}}{\omega_n^{(\textsc{AE})}\omega_n^{(\textsc{EB})}-\rho_{\textsc{AB}}^2},
    \label{eq:Cn}
\end{equation}
\begin{equation}
D_n = \frac{\rho_{\textsc{AE}}\omega_n^{(\textsc{AE})} - \rho_{\textsc{AB}}\rho_{\textsc{EB}}}{\omega_n^{(\textsc{AE})}\omega_n^{(\textsc{EB})}-\rho_{\textsc{AB}}^2},
    \label{eq:Dn}
\end{equation}
\end{subequations}
where $\omega_n^{(\textsc{AE})} = 1 + \frac{\sigma_{\textsc{AE}}^2}{\lambda_n}$, $\omega_n^{(\textsc{EB})} = 1 + \frac{\sigma_{\textsc{EB}}^2}{\lambda_n}$. According to the previous definition, $\rho_{\textsc{AB}}$ represents the spatial correlation between the two channels observed by Eve, i.e. the Alice-Eve channel and the Eve-Bob one.
$\lambda_n$ is the power delay, and we suppose that this value is always known to the attacker, thus considering a worst case scenario, in which Eve is in an advantageous condition to perform her attack.
The case of Eve not knowing exactly the power delay profile is more realistic, but puts Eve in a less favorable situation and is out of the scope of this paper.

\subsection{Attack to the combined test}

In order to find the optimal attack strategy for Eve when Bob uses the combined test described in Section \ref{sec:CombinedTest}, denoted as \textit{modulus attack} for brevity, we consider the worst case scenario, where Eve is able to perfectly estimate $\mathbf{h}^{\textsc{(AE)}}$ and $\mathbf{h}^{\textsc{(EB)}}$, i.e. $\sigma_{\textsc{AE}}^2 = \sigma_{\textsc{EB}}^2 = 0$. For the sake of simplicity we also suppose that no correlation exists between the two channels observed by her or, in other words, $\rho_{\textsc{AB}} \rightarrow 0$.

Under these hypotheses \eqref{eq:att1} boils down to
\begin{equation}
    g_n =  \rho_{\textsc{AE}}\hat{h}_{\textsc{AE}} + \rho_{\textsc{EB}}\hat{h}_{\textsc{EB}}.
    \label{eq:att_simpl}
\end{equation}

On the other hand, in order to make $\left|\mathbf{g}\right|$ more similar to $\left|\hat{\mathbf{h}}^{(\textsc{AB})} \right|$, in the modulus attack Eve can choose 
\begin{equation}
    g_n = \frac{\hat{h}_n^{(\textsc{AE})}}{\rho_{\textsc{AE}}} + \frac{\hat{h}_n^{(\textsc{EB})}}{\rho_{\textsc{EB}}}, \mbox{ for } \rho_{\textsc{AE}}, \rho_{\textsc{EB}}\neq 0. 
\label{eq:att_mod}
\end{equation}

When we consider a decision strategy based on both \ac{LLR} and modulus comparison, however, attacks based on \eqref{eq:att_simpl} and \eqref{eq:att_mod} are no longer optimal. 
In this case, the best attack strategy in fact consists in forging
\begin{equation}
g_n = \rho_{\textsc{\textsc{AE}}}^x \hat{h}_n^{(\textsc{AE})} + \rho_{\textsc{\textsc{EB}}}^y \hat{h}_n^{(\textsc{EB})},
    \label{eq:att3}
\end{equation}
where $x,y \in [-1,1]$. 
This requires finding a trade-off between the modulus attack and the \ac{LLR} attack. In fact, $x = y = -1$ corresponds to the best attack to the modulus comparison method, while $x =y = 1$ corresponds to the best attack strategy to the \ac{LLR}.
Finding the optimal values of the couple $(x,y)$ corresponds to solve the problem 

\begin{equation}
     (x,y) = \arg \max_{x,y \in [-1,1]} P\left[\Psi(x) \leq \theta, \Gamma(x)^2 \leq \epsilon^2 | \mathbf{g} = \eqref{eq:att3} \right],
    \label{eq:opt}
\end{equation}
i.e. to find the values of $(x,y)$ that give the highest $P_{\textsc{MD}}$.

In Table \ref{tab:x_val} we report the optimal values of $(x,y)$ obtained by solving \eqref{eq:opt} through numerical methods, with $P_{\textsc{FA}} = 10^{\textsc{-4}}$ and different values of sub-carriers $N$ and the time-varying fading $\boldsymbol{\alpha}^{\textsc{(II)}}$, considering the absence of time-varying fading in Phase 1 ($\boldsymbol{\alpha}^{\textsc{(I)}} = 1$). We do not report values corresponding to $\rho_{\textsc{AE)}} \geq 0.7$, which are always equal to 1.
We observe that for high values of the spatial correlation the attack to the \ac{LLR} test results to be most convenient strategy to adopt for Eve even when Bob chooses the combined test.
The case with $\rho_{\textsc{AE}} = 1$ is of special interest: Eve is exactly is Alice's same position, so each kind of attack results successful regardless of the choice of $(x,y)$. 
Since the attacker does not know exactly the values of $\boldsymbol{\alpha}^{\textsc{(II)}}$, her most conservative choice is to suppose that all the entries of $\boldsymbol{\alpha}^{\textsc{(II)}}$ are equal to $1$. 
In fact, Eve is in the worst condition in absence of fading, since Bob lets fewer messages to be accepted as authentic.

\begin{table}[t]
\begin{center}
\caption{Optimal values of the exponents $(x,y)$ in Eq. (\ref{eq:att3}) for different values of $\boldsymbol{\alpha}^{\textsc{(II)}}$ and $N$, with $\boldsymbol{\alpha}^{\textsc{(I)}} = 1$ and $\rho_{\textsc{AE}} = \rho_{\textsc{EB}} = \rho $. \label{tab:x_val}
}
\begin{tabular}{|@{\hspace{1mm}}c@{\hspace{1mm}}|@{\hspace{1mm}}c@{\hspace{1mm}}|@{\hspace{0.5mm}}c@{\hspace{0.5mm}}|@{\hspace{0.5mm}}c@{\hspace{0.5mm}}|@{\hspace{0.5mm}}c@{\hspace{0.5mm}}|@{\hspace{0.5mm}}c@{\hspace{0.5mm}}|@{\hspace{0.5mm}}c@{\hspace{0.5mm}}|@{\hspace{0.5mm}}c@{\hspace{0.5mm}}|}
\hline
\multirow{2}*{$\boldsymbol{\alpha}^{\textsc{(II)}}$} & \multirow{2}*{$N$} & \multicolumn{6}{c|}{$\rho$} \\
\cline{3-8}
 & & 0.1 & 0.2 & 0.3 & 0.4 & 0.5 & 0.6 \\
\hline
\multirow{3}*{1} & $1$ & $(0.7,0.7)$ & $(0.8,0.8)$ & $(0.9,1)$ & $(1,1)$ & $(1,1)$ & $(1,1)$  \\
\cline{2-8} 
& $3$ & $(0.7,0.7)$ & $(0.8,0.9)$ & $(1,1)$ & $(1,1)$ & $(1,1)$ & $(1,1)$   \\
\cline{2-8}
& $6$ & $(0.8,0.6)$ & $(0.9,0.9)$ & $(1,1)$ & $(1,1)$ & $(1,1)$ & $(1,1)$   \\
\hline
\multirow{3}*{0.8} & 1 & $(0.4,0.5)$ & $(0.5,0.6)$ & $(0.6,0.6)$ & $(0.7,0.6)$ & $(0.8,0.8)$ & $(0.9,1)$  \\
\cline{2-8} 
& $3$ & $(0.4,0.5)$ & $(0.5,0.6)$ & $(0.6,0.6)$ & $(0.8,0.6)$ & $(0.8,0.8)$ & $(1,1)$  \\
\cline{2-8}
& $6$ & $(0.5,0.3)$ & $(0.6,0.4)$ & $(0.5,0.6)$ & $(0.7,0.6)$ & $(0.7,0.8)$ & $(0.9,0.9)$  \\
\hline
\end{tabular}
\end{center}
\end{table}

\subsection{Mismatched Attacker}

Let us consider an attacker who does not exactly know which is the decision criterion used by Bob. In such a case, Eve has no option but to try to design $\mathbf{g}$ such as it gives the best probability of \ac{MD} in each case. In particular, Eve's problem here is to decide a general strategy which leads to the highest possible probability of missed detection whatever is the decision method adopted by Bob.

\begin{figure}[!t]
\begin{centering}
 \subfigure[]
   {\includegraphics[width=0.45\textwidth]{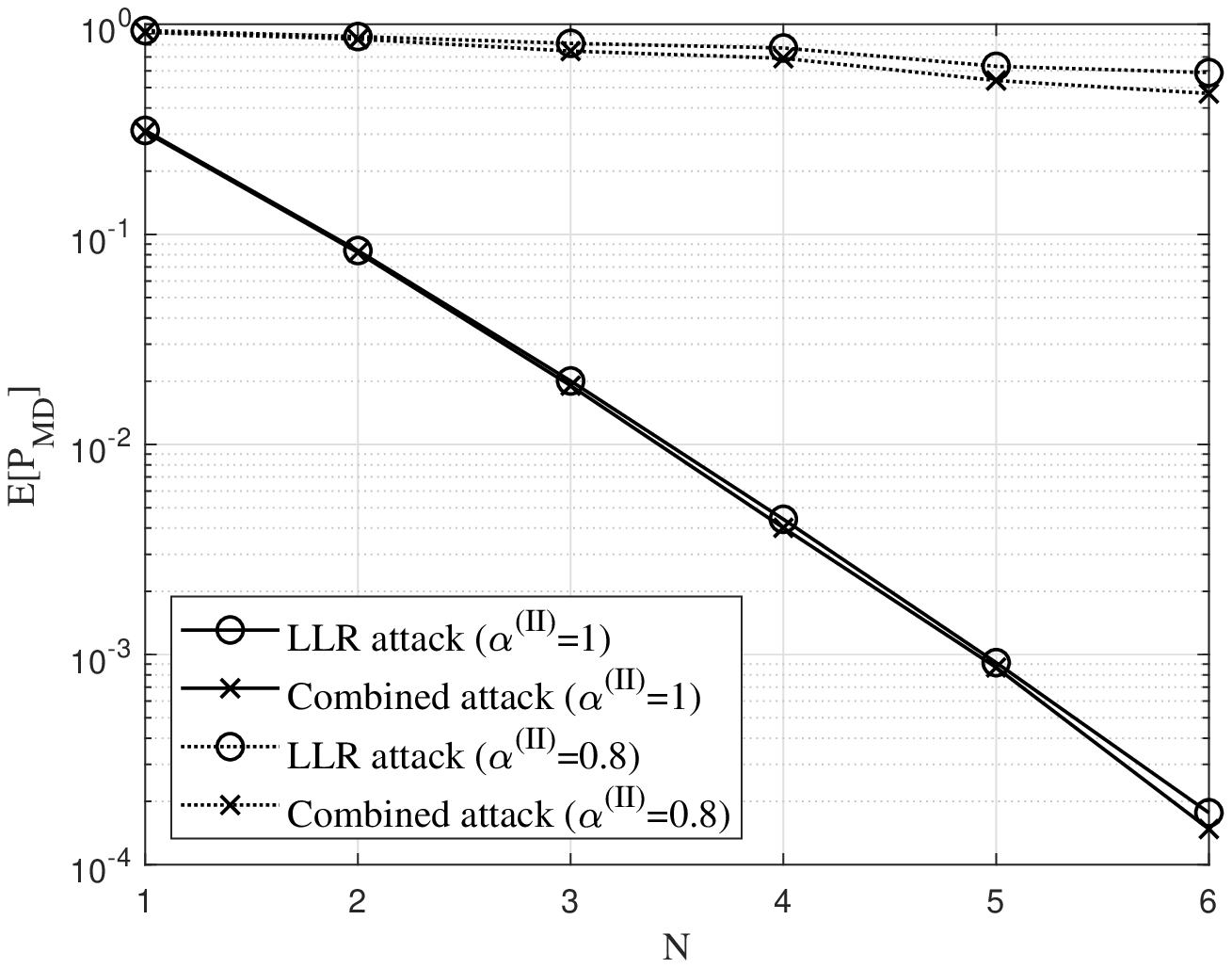}
   \label{fig:mis_comb_a}}
 \subfigure[]
   {\includegraphics[width=0.45\textwidth]{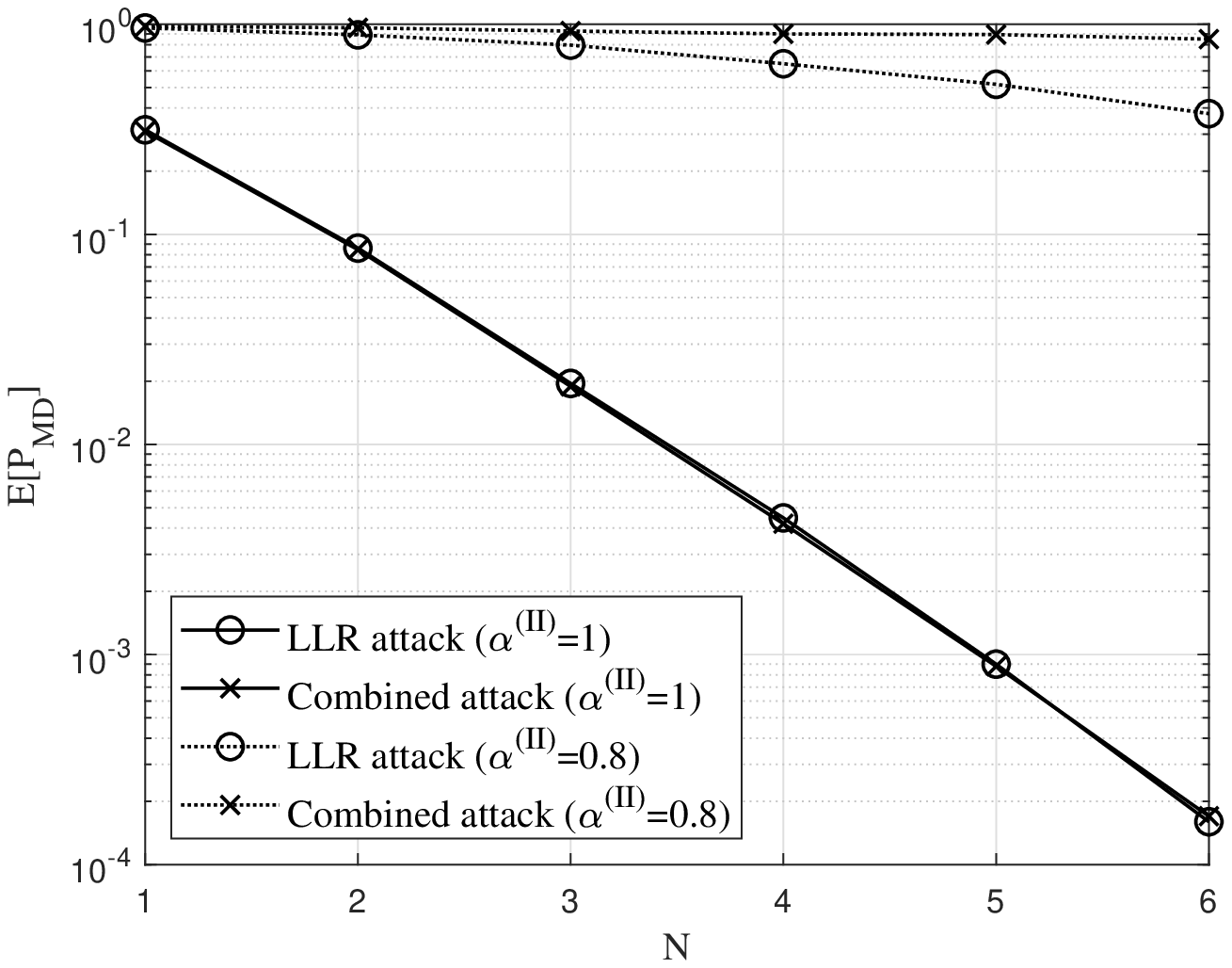}
   \label{fig:mis_comb_b}}
\caption{Performance of the matched versus mismatched attacker in terms of average $P_{\textsc{MD}}$, when Bob applies the a) LLR test and b) combined test, with $\boldsymbol{\alpha}^{\textsc{(I)}} = 1$, $\rho_{\textsc{AE}} = \rho_{\textsc{EB}} = 0.1$.}
\label{fig:mis_comb}
\par\end{centering}
\end{figure}

In Figs. \ref{fig:mis_comb_a} and \ref{fig:mis_comb_b} we compare the average probability of \ac{MD} obtained when Eve's attack strategy follows \eqref{eq:att_simpl}, \eqref{eq:att_mod} and \eqref{eq:att3} with
the values of $(x,y)$ found in Table \ref{tab:x_val} for $\rho_{\textsc{AE}} = \rho_{\textsc{EB}} = 0.1$ and $\boldsymbol{\alpha}^{\textsc{(II)}} = [0.8, 1]$ on each sub-carrier. It is possible to note that using the wrong attack strategy is penalizing especially when Bob exploits a combined test and when $\boldsymbol{\alpha}^{\textsc{(II)}} < 1$, and this is particularly evident for a large value of $N$. In case of stationary channels, i.e. $\boldsymbol{\alpha}^{\textsc{(II)}} = 1$, performances are almost equivalent for both decision strategies.

\section{Methods based on machine learning}
\label{sec:machine_learning}

Let us now consider the application of machine learning techniques to the authentication protocol presented in Sec. \ref{sec:sys_model}. These methods have the advantage of not requiring any knowledge about the receiver's \ac{CSI}, not even the variance $\sigma^2$ needed by the statistical tests illustrated in Sec. \ref{sec:stat_methods}. They are hence able to perform authentication in a blind way.

We consider two different scenarios. In the ideal one, the authenticator has some information about the attacker, thus Phase I corresponds to the training phase of a binary classifier. In the second one, more realistic, Bob is able to collect only samples belonging to the legitimate transmitter during Phase I and we must consider \ac{OCC}, where only one class (referred to as positive class or target class) is present during the training phase, while the others (negative classes or non-target classes) are not known \cite{khan2009}.

\acs{OCC} methods are mainly based on two parameters \cite{Tax2001}: the distance $d(x)$ between the sample to classify and the target class, and the threshold $\theta_d$ on the value of the distance.
Formally, a new instance is recognized as belonging to the positive class if its distance from the target class is below the threshold. In formulas
\begin{equation}
    f_{oc}(x) = I\left(d(x) < \theta_{d}\right)
\label{eq:oneclass}
\end{equation}
where $I(\cdot)$ represents an indicator function\footnote{An indicator function is a function defined on a set \textit{X} that indicates membership of an element in a subset \textit{A} of \textit{X}. Its value is 1 for all elements of \textit{A} and 0 for all elements of \textit{X} not in \textit{A}.}
and $f_{oc}(x)$ is the decision function, i.e. a binary function expressing acceptance of the object $x$ into the target class. 
There are several metrics which can be used to evaluate $d(x)$, among which the most common choice is the Euclidean distance. 

\acs{OCC} approaches differ in the evaluation of the distance and in the optimization of the threshold, on the basis of the available training set. 
In the following, we consider two different algorithms, i.e. the \ac{NN} and the \ac{SVM} techniques.

\subsection{Nearest Neighbors}
\label{sec:ocnn}

\ac{NN} algorithms classify a sample $x$ by assigning it the most frequent label among its $k$ nearest neighbors. 
What makes \ac{NN} algorithms particularly suitable for an authentication scheme to be implemented in resource-constrained devices is their low complexity, which grows with the training set dimension. 
They operate classification only on instances of the training set and not on statistical assumptions. 
Since in authentication schemes it is reasonable to assume that phase I is short and Bob receives a limited number of setup packets, it is also reasonable to assume to work with a small training set, thus making the choice of classification based on \ac{NN} algorithms a good trade-off between complexity and performance.

\ac{OCNN} techniques are derived from traditional binary \ac{NN} algorithms. In detail, a \ac{OCNN} algorithm works as follows \cite{Khan2018}: it finds the \textit{j} nearest neighbors $\left\{y_1, \cdots, y_j \right\}$ of the test sample $x$ in the target class, and the \textit{k} nearest neighbors $\left\{z_{i1}, \cdots, z_{ik} \right\}$ of the first \textit{j} neighbors; it evaluates the average distance $\bar{D}_{xy}$ over $\left\{D_{xy_1}, \cdots, D_{xy_j} \right\}$ and the average distance $\bar{D}_{yz}$ over $\left\{D_{{y_1}{z_{11}}}, \cdots, D_{{y_1}{z_{1k}}}, \cdots, D_{{y_1}{z_{k1}}}, \cdots, D_{{y_k}{z_{kk}}} \right\}$; $x$ is then considered as a member of the target class if
    \begin{equation}
    \frac{\bar{D}_{xy}}{\bar{D}_{yz}} < \theta_d.
    \label{eq:JKNN}
    \end{equation}
\ac{OCNN} methods can be grouped into four main categories (11NN, 1KNN, J1NN, JKNN), depending on which of the parameters $j$ and $k$ is fixed to 1. They differ in the number of neighbors used to compute the decision threshold \cite{Khan2018}. 

When the training set is not affected by time-varying fading while the test set is, the probability of FA increases with the number of features. As an example, let us consider a 11NN algorithm, which accepts a new packet if \eqref{eq:JKNN} is satisfied. $y$ and $z$ both belong to the training set, thus $D_{yz}$ is not influenced by the time-varying fading, differently from $D_{xy}$. Both distances increase with the number of features $2N$, whether they are computed using a Euclidean distance or the \ac{LLR} considered in Section \ref{subsec:hybridapproach}.
In the stationary case, the increase in $D_{xy}$ following from an increased number of features is balanced by an increase in $D_{yz}$, and their ratio does not increase with the number of features. The presence of time-varying fading instead makes $D_{xy}$ increase more than in the stationary case, and in this case such an effect is only partly counterbalanced by an increase in $D_{yz}$.
The threshold $\theta_d$ remains the same in both cases, being computed only on the basis of the training phase, which does not include time-varying fading.
Therefore, by increasing the number of features, it is more likely that the ratio $D_{xy}/D_{yz}$ overcomes $\theta_d$, causing the rejection of an authentic message and the occurrence of a FA. 
These considerations can be easily extended to all the \ac{OCNN} methods examined in this paper.

\subsection{Support Vector Machines}

Despite their low complexity, \ac{NN} algorithms require to determine some optimal parameters, such as $j$, $k$ and $\theta_d$, which can become computationally expensive. For this reason, we also consider a second kind of classifiers, known as \ac{SVM}.

In general, \acp{SVM} can create a non-linear decision boundary to separate different classes by projecting the data through a non-linear function to a space with a higher dimension, lifting them from their original space to a feature space, which can be of unlimited dimension. 
Their one-class version, also known as single-class classification or \textit{novelty detection}, was introduced in \cite{Scholkopf1999}. 
The main concept behind the \acs{OC-SVM} algorithm consists in obtaining a spherical boundary, in feature space, around the data. The volume of this hyper sphere is minimized, to minimize the effect of incorporating outliers in the solution \cite{Tax2004}.
In particular, the goal of \acs{OC-SVM} is to estimate a function $f_{oc}(x)$ that encloses the most of training data into a hyper sphere $R_x = \left\{x \in R^N | f_{oc}(x) > 0 \right\}$, where $N$ is the size of feature vector. The decision function $f_{oc}(x)$ is written as
\begin{equation}
    f_{oc}(x) = sgn\left\{\sum_{i=1}^m \lambda_i K(x,x_i)-\xi \right\},
    \label{eq:svm}
\end{equation}
where $m$ represents the number of training samples, $\xi$ is the distance of the hyper sphere from the origin and $K(\cdot, \cdot)$ defines
the \acs{OC-SVM} kernel that allows projecting data from the original space to the feature space. $\lambda_i$ are the Lagrange multipliers computed by optimizing the following equations
\begin{equation}
    \min_{\lambda} \left\{ \frac{1}{2} \sum_{i,j} \lambda_i \lambda_j K(x_i,x_j) \right\},
\end{equation}
subject to $0 \leq \lambda_i \leq \frac{1}{\nu m}$ and $\sum_{i=1}^m \lambda_i = 1$,
where $\nu$ is the percentage of data considered as outliers. 
Modifying the parameter $\nu$ allows to optimize the value of $g_{mean}$ or to minimize one of the two probabilities of FA and MD.

A pattern $x$ is accepted if $f_{oc}(x) > 0$ and rejected otherwise. Different functions can be used, such as linear, polynomial or Gaussian kernels. Usually, the Gaussian is the most used kernel, which allows determining the radius of the hyper sphere according the parameter $1/2\sigma_{\textsc{SVM}}^2$. It is defined as
\begin{equation}
    K(x,x_i) = \exp\left(-\frac{ ||x-x_i||^2}{2\sigma_{\textsc{SVM}}^2}\right),
\end{equation}
where the numerator represents the squared Euclidean distance between two general feature vectors, while $\sigma_{\textsc{SVM}}$ is a free parameter.

\subsection{Clustering}

All previous classification mechanisms require that Bob exactly knows the source of one or more packets during the training phase.
This assumption requires the use of higher layer cryptographic protocols or manual solutions, which may result unpractical in some cases.
To address this issue, we also consider the use of clustering algorithms \cite{Jain1999} for Bob to establish the origin of packets during the training phase and to assign them a possibly correct label.
In particular, we focus on the $k$-means approach \cite{MacQueen1967}, which guarantees good performance and a fast convergence; moreover, with a small input set and when the number of clusters $k$ is known and small, it also has a reduced complexity.

Given a set of $m$ instances $({x}_1, {x}_2, \cdots, {x}_m)$, the goal of $k$-means clustering is to partition them into $k \leq m$ sets (or \textit{clusters}) $\mathbf{S} = \left\{S_1, S_2, \cdots, S_k\right\}$ such as to minimize the within-cluster sum of squares (WCSS). More formally, the objective is to find
\begin{equation}
    \arg \min_{S} \sum_{i=1}^k \sum_{x \in \mathcal S_i} \|x-{\mu}_i \|^2 ,
    \label{eq:kmeans}
\end{equation}
where $\mu_i$ is the mean of points in $S_i$.
Classes labels in the training phase will be determined by results got from the algorithm and not assigned a priori by the authenticator.

\subsection{Hybrid approach \label{subsec:hybridapproach}}

Let us consider a hybrid approach that exploits the statistical metrics considered in Section \ref{sec:llr} along with \ac{OCNN} classifiers described in Section \ref{sec:ocnn}.
This is done by using in \eqref{eq:JKNN} the \ac{LLR} defined in \eqref{eq:psi} instead of the Euclidean distance.
With respect to the Euclidean distance, computing the \ac{LLR} requires knowledge of the statistical distribution through the value of $\sigma^2$. 
When the \ac{LLR} is used along with a JKNN classifier, we select the $j$ samples from the training set that minimize the \ac{LLR} evaluated on the new sample to classify, and the $k$ samples that minimize the \ac{LLR} computed on the $j$ samples.
Analogously, the \ac{LLR} can be used with all the considered \ac{NN} methods.

\section{Results and discussion}
\label{sec:results}

In this section, we assess and compare the performance of statistical and machine learning-based decision methods under different system conditions and assumptions. For the sake of simplicity, and without loss of generality, we consider examples in which time-varying fading affects all channels in the same way during both authentication phases, i.e. $\alpha_n = \alpha$ for $n = 1, \cdots, N$. We also suppose that Eve can estimate the channel between Alice and herself, but not the channel between her and Bob.
Bob in fact receives messages from Alice but is not expected to transmit enough messages to allow the attacker to extract useful information on $\mathbf{h}^{(\textsc{EB})}$, thus imposing $\rho_{\textsc{AB}} \rightarrow 0$, $\rho_{\textsc{EB}} \rightarrow 0$, $\rho_{\textsc{AE}} > 0$. Moreover, we assume to give Eve the maximum advantage, allowing her to perfectly estimate $\mathbf{h}^{\textsc{(AE)}}$, i.e. considering $\sigma_{\textsc{AE}}^2 = 0$.
The average \ac{SNR} on channel estimates during both phases is $\textsc{SNR}^{(\textsc{I})} = 1/\sigma_{\textsc{I}}^2$ and $\textsc{SNR}^{(\textsc{II})} = 1/\sigma_{\textsc{II}}^2$.

We compare the performance of statistical and machine learning-based decision methods by giving them the same inputs.
To this end, the features we use for machine learning are represented by the real and the imaginary parts of the channel coefficient measured on each sub-carrier for both \ac{NN} and \ac{SVM} algorithms. Therefore the number of features corresponds to $2N$.

\subsection{Parameter setting and \ac{OCC} performance evaluation}

We first examine the security performance achievable by \ac{OCC} techniques. 
As already said in Sec. \ref{sec:machine_learning}, the definition of the parameters of \ac{NN} algorithms has an important impact on the precision of the results. In order to find a trade-off between false alarm and missed detection, the parameters $j$, $k$ and $\theta_d$ have been optimized in such a way as to maximize $g_{mean}$, by using a $g$-fold cross validation. As regards binary $k$-NN, common values for $k$ are usually in the range $[3, \sqrt{M}]$, given $M$ as the number of samples in the training set \cite{Duda2000}, and $k$ must be an odd number in order to avoid parity issues. 
As an example, the results obtained for $\alpha^{\textsc{(I)}} = \alpha^{\textsc{(II)}} = 1$ and $N = 3$ are reported in Tab. \ref{tab:parametri1}, choosing $g$ equal to 5 and considering two training sets of $100$ and $1000$ samples (with $50\%$ of samples belonging to the positive class and $50\%$ to the negative class for the $k$-NN).

\begin{figure}[ht]
\centering
\includegraphics[width=0.45\textwidth]{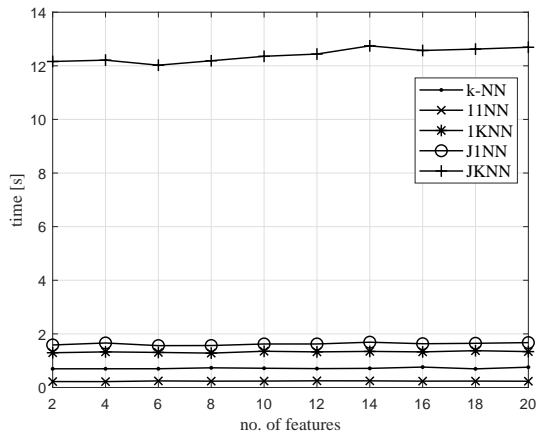}
\caption{Training times required for different NN algorithms with  $\alpha^{\textsc{(I)}} = 0.8$, $\alpha^{\textsc{(II)}} = 0.9$,  $\textsc{SNR}^{(\textsc{I})} = 15$dB, $\textsc{SNR}^{(\textsc{II})} = 20$dB and $M = 100$.  \label{fig:train_times}}
\end{figure}

\ac{OCNN} algorithms have been implemented in Matlab 2019a. The presented case studies were run on a machine equipped with an Intel Core i7-8565U 3.9 Ghz Quad-core processor and 16 GB of RAM.  The number of features apparently has an almost irrelevant impact on the computational times, as shown in Fig. \ref{fig:train_times}. 
In this kind of classifiers, in fact, the number of features, i.e. the length of the input vectors, only affects the initial computation of the distance (in our case the Euclidean distance) between samples. For the relatively small numbers of features considered in Fig. \ref{fig:train_times}, the variation in the computation time of the Euclidean distance for vectors of increasing lengths is almost negligible and has a small effect on the training time. We note that, as predictable, JKNN requires a larger time than the other methods, especially compared to 11NN, which needs only a threshold optimization.

\begin{table}[ht]
\begin{center}
\caption{Optimized parameters for NN methods, with $\alpha^{\textsc{(I)}} = \alpha^{\textsc{(II)}} = 1$ and $N = 3$, considering two training sets of 100 and 1000 samples (into brackets). \label{tab:parametri1}}
\begin{tabular}{|@{\hspace{1mm}}c@{\hspace{1mm}}|@{\hspace{1mm}}c@{\hspace{1mm}}|@{\hspace{1mm}}c@{\hspace{1mm}}|@{\hspace{1mm}}c@{\hspace{1mm}}|@{\hspace{1mm}}c@{\hspace{1mm}}|@{\hspace{1mm}}c@{\hspace{1mm}}|@{\hspace{1mm}}c@{\hspace{1mm}}|}
\hline
& & KNN & 11NN & 1KNN & J1NN & JKNN   \\
\hline
\hline
\multirow{3}*{$\rho_{\textsc{AE}} = 0.1$} & $k$ & 9 (31) & - & 3 (3) & - & 2 (2) \\
\cline{2-7}
& $j$ & - & - & - & 2 (2.5) & 2 (2)  \\
\cline{2-7}
& $\theta_d$ & - & 3 (3.5) & 2.5 (3) & 3 (3.5) & 2.5 (2.5) \\
\hline
\multirow{3}*{$\rho_{\textsc{AE}} = 0.8$} & $k$ & 3 (7) & - & 5 (27) & - & 6 (22) \\
\cline{2-7}
& $j$ & - & - & - & 7 (25) & 6 (22)  \\
\cline{2-7}
& $\theta_d$ & - & 2 (2) & 1.5 (1) & 2 (2.5) & 2 (1.5) \\
\hline
\end{tabular}
\end{center}
\end{table}

As for \ac{SVM} algorithms, we exploit a modified version of the tool in \cite{scikit-learn}. We use a Gaussian kernel, which proved to be more effective than linear and polynomial kernels in our scenario, as it is evident from the results in Tab. \ref{tab:kernel}.

Results achieved by applying different \ac{OCC} algorithms to the same data set are then compared. In the following examples, we consider $\textsc{SNR}^{(\textsc{I})} = 15$dB and $\textsc{SNR}^{(\textsc{II})} = 20$dB, fading coefficients $\alpha^{\textsc{(I)}} = 1$ and $\alpha^{\textsc{(II)}} = 0.9$, a training set of $M = 100$ samples and a classification set composed by 5 subsets of $4\cdot10^5$ elements. Each subset contains 50$\%$ of positive samples and $50\%$ of negative samples.
The values of $P_{\textsc{MD}}$ and $P_{\textsc{FA}}$ are averaged over 100 different datasets randomly generated.

\begin{table}
\begin{center}
\caption{Results obtained by SVM with different kernels, with $M=100$, $\rho_{\textsc{AE}} = 0.8$, $\alpha^{\textsc{(I)}} = \alpha^{\textsc{(II)}} = 1$ and a dataset dimension equal to 1000.
\label{tab:kernel}}
\begin{tabular}{|@{\hspace{0.3mm}}c@{\hspace{0.3mm}}|@{\hspace{0.3mm}}c@{\hspace{0.3mm}}|@{\hspace{0.3mm}}c@{\hspace{0.3mm}}|@{\hspace{0.3mm}}c@{\hspace{0.3mm}}|@{\hspace{0.3mm}}c@{\hspace{0.3mm}}|@{\hspace{0.3mm}}c@{\hspace{0.3mm}}|@{\hspace{0.3mm}}c@{\hspace{0.3mm}}|@{\hspace{0.3mm}}c@{\hspace{0.3mm}}|}
\hline
& $N$ & 1 & 2 & 3 & 4 & 5 & 6 \\
\hline
\hline
\multirow{2}*{Gaussian} & $P_{\textsc{FA}}$ & 0.001 & $<0.001$ & $<0.001$ & $<0.001$ & $<0.001$ & $<0.001$  \\
& $P_{\textsc{MD}}$ & 0.08 & 0.024 & 0.05 & 0.01 & 0.001 & $<0.001$ \\
\hline
\multirow{2}*{Poly} & $P_{\textsc{FA}}$ & 0.019 & 0.007 & 0.022 & 0.002 & 0.03 & 0.005  \\
& $P_{\textsc{MD}}$ & 0.119 & 0.064 & 0.058 & 0.04 & 0.02 & 0.017 \\
\hline
\multirow{2}*{Linear} & $P_{\textsc{FA}}$ & 0.019 & 0.007 & 0.02 & 0.002 & 0.029 & 0.005  \\
& $P_{\textsc{MD}}$ & 0.118 & 0.064 & 0.058 & 0.04 & 0.02 & 0.017 \\
\hline
\end{tabular}
\end{center}
\end{table}

\begin{figure}[ht]
\centering
\includegraphics[width=0.45\textwidth]{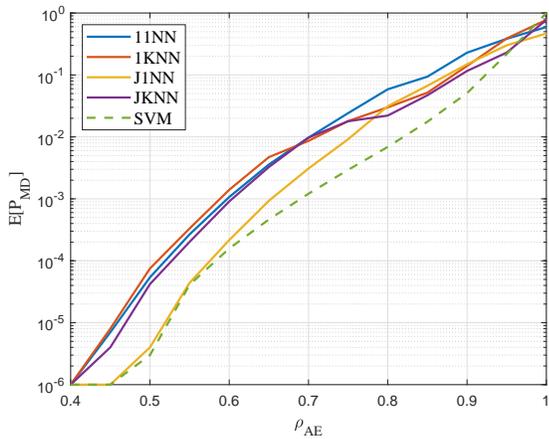}
\caption{
Probability of MD as a function of the spatial correlation $\rho_{\textsc{AE}}$, comparing NN and SVM algorithms, with $\alpha^{\textsc{(I)}} = 1$, $\alpha^{\textsc{(II)}} = 0.9$, $N = 3$,  $\textsc{SNR}^{(\textsc{I})} = 15$dB, $\textsc{SNR}^{(\textsc{II})} = 20$dB and $M = 100$ (training set dimension).  \label{fig:rhoVSpmd}}
\end{figure}

\begin{figure}[!t]
\begin{centering}
 \subfigure[]
   {\includegraphics[width=0.45\textwidth]{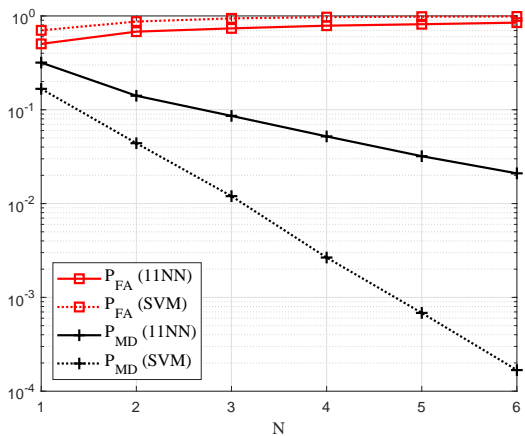}
   \label{fig:confronto_a}}
 \subfigure[]
   {\includegraphics[width=0.45\textwidth]{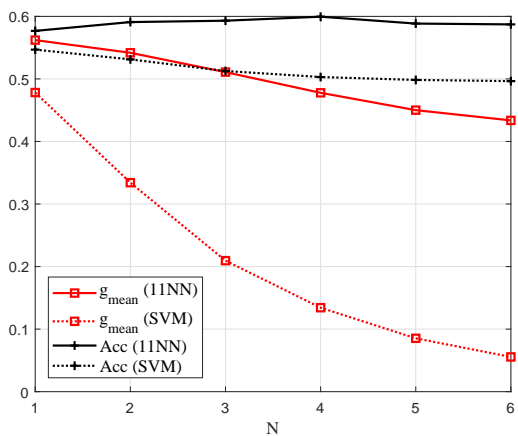}
   \label{fig:confronto_b}}
\caption{
a) Probabilities of FA and MD and b) $g_{\textsc{mean}}$ and $\textsc{Acc}$ as function of the number of sub-carriers $N$, comparing 11NN and SVM algorithms, with $\alpha^{\textsc{(I)}} = 1$ and $\alpha^{\textsc{(II)}} = 0.9$, $\rho_{\textsc{AE}} = 0.8$, $\textsc{SNR}^{(\textsc{I})} = 15$dB, $\textsc{SNR}^{(\textsc{II})} = 20$dB and $M=300$. }
\label{fig:confronto}
\par\end{centering}
\end{figure}

In Fig. \ref{fig:rhoVSpmd} the performance obtained for different values of the spatial correlation coefficient $\rho_{\textsc{AE}}$ is shown, with the meaning of Eve
being at decreasing distances from Bob, while in Fig. \ref{fig:confronto}a results are obtained for different numbers of sub-carriers.
We can note that, in general, there is no \ac{OCNN} algorithm that results to be the best one for any value of $\rho_{\textsc{AE}}$. However, for a given similar performance, 11NN requires much smaller training and classification times, especially in comparison with JKNN. Under the considered assumptions, \ac{SVM} obtains the lowest probability of \ac{MD}, but at the expenses of a higher \ac{FA}. The \ac{SVM} in fact obtains a $P_{\textsc{FA}}$ equal to 0.944, while for J1NN it is equal to 0.716 (similar values are obtained by the other NN algorithms and not reported here for brevity). For both the examined methods we consider the parameters, i.e. the values of $j$, $k$ and $\theta_d$ for the \ac{OCNN} classifiers and $\nu$ as regards the \ac{SVM}, that achieve the best value of $g_{mean}$ during the training phase. By looking at the curves it is possible to observe that until $\rho_{\textsc{AE}} \leq 0.4$ all the algorithms exhibit a \ac{MD} probability lower than $10^{-6}$, meaning that no \ac{MD} event has been observed over the entire classification set, of dimension $10^6$.
Moreover, as expected, Fig. \ref{fig:confronto_a} testifies the fact that an increasing number of features is beneficial from a security point of view, even if the probability of FA slightly increases with the number of features, as discussed in Section \ref{sec:ocnn}. According to the considerations reported at the end of Section \ref{sec:ocnn}, the presence of fading, by means of the parameter $\alpha^{\textsc{(II)}}$, negatively influences the probability of false alarm achieved by \ac{OCNN} classifiers with an increasing number of subcarriers, and therefore of features. It is reasonable to suppose that a similar behavior also affects the \ac{SVM}, leading to the reduction of performance shown in Fig. \ref{fig:confronto_a}. The worsening of the false alarm probability, together with an improvement of the missed detection probability, inevitably deteriorates the performance in terms of $g_{mean}$, which gives a measure of the balance between the values of $P_{\textsc{FA}}$ and $P_{\textsc{MD}}$ (see Eq. \eqref{eq:gmean}). Fig. \ref{fig:confronto_b} reports the optimal values of $g_{mean}$ for increasing values of $N$, showing that \ac{NN} algorithms are able to achieve a better $g_{mean}$ than \acp{SVM}.
For the sake of comparison, we also show the performance in terms of accuracy, which exhibits less sensitivity to the number of features with respect to $g_{mean}$.

\subsection{Impact of the training phase}

The training phase, or Phase I, has clearly a relevant impact on the subsequent classification phase performance. Let us assess how different training parameters and conditions influence the system performance.

\begin{figure}[t]
\centering
\includegraphics[width=0.45\textwidth]{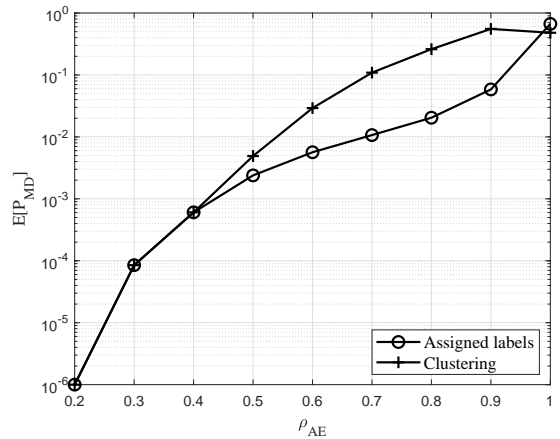}
\caption{
Average probability of MD versus $\rho_{\textsc{AE}}$ using clustering with respect to known instance labels, with $N = 6$, $M=200$, $\alpha^{\textsc{(I)}} = \alpha^{\textsc{(II)}} = 1$, $\textsc{SNR}^{(\textsc{I})} = 15$dB, $\textsc{SNR}^{(\textsc{II})} = 20$dB, classification performed using a SVM.}
\label{fig:cluster}
\end{figure}

Firstly, reliability of the training set must be taken into account.
In the ideal case, instances labels forming the training set are exactly known.
In real cases, instead, they may be unknown, and must be hence inferred through clustering techniques.
A comparison of these two cases in terms of probability of \ac{MD} is illustrated in Fig. \ref{fig:cluster}, where a training procedure based on clustering, and in particular on k-means algorithm, is compared with one in which instance labels are exactly known.
In both cases, a two-class \ac{SVM} algorithm with Gaussian kernel has been applied over a training set of 200 samples, 100 from Alice and 100 from Eve. k-means requires as initial choice the knowledge of the number of cluster, which in our case is known by hypothesis and is equal to 2, and the initial partitions. Initial cluster centers, or \textit{centroids}, have been chosen by selecting observations uniformly at random from the dataset. Different initial centroids, however, may produce different final clusters. In order to address this problem, results have been averaged over 50 possible different initial choices.
Despite being disadvantaged, clustering methods are able to achieve the same performance of methods with assigned labels for low values of $\rho_{\textsc{AE}}$ but, as expected, they achieve larger $P_{MD}$ as soon as $\rho_{\textsc{AE}}$ increases. The probability of \ac{FA} remains fixed and lower than $10^{-6}$, except for $\rho_{\textsc{AE}} = 1$. 

In Tab. \ref{tab:train_set} we compare the results obtained with training sets of different dimension by using a 11NN (chosen among the other \ac{OCNN} methods thanks to its low complexity, which makes it able to train a set of dimension 10,000 in reasonable times) and a \ac{SVM} algorithm. In both cases, we observe that the value of $M$ has a small impact on the results, and the only relevant improvement is given on the $P_{\textsc{FA}}$ of the \ac{SVM} for $\alpha^{\textsc{(II)}} = 1$ . This implies that we can use a small training set without incurring in any significant performance degradation.

\begin{table}
\begin{center}
\caption{Results for different training set dimensions choosing $N = 3$, $\rho_{\textsc{AE}} = 0.8$ and $\alpha^{\textsc{(I)}} = 1$. \label{tab:train_set}}
\begin{tabular}{|@{\hspace{0.3mm}}c@{\hspace{0.3mm}}|@{\hspace{0.3mm}}c@{\hspace{0.3mm}}|@{\hspace{0.3mm}}c@{\hspace{0.3mm}}|@{\hspace{0.3mm}}c@{\hspace{0.3mm}}|@{\hspace{0.3mm}}c@{\hspace{0.3mm}}|@{\hspace{0.3mm}}c@{\hspace{0.3mm}}|@{\hspace{0.3mm}}c@{\hspace{0.3mm}}|}
\hline
& & dim & 10 & 100 & 1,000 & 10,000 \\
\hline
\hline
\multirow{4}*{$\alpha^{\textsc{(II)}} = 1$} & \multirow{2}*{SVM} &  $P_{\textsc{FA}}$ & $7.4\cdot10^{-3}$ & $1.93\cdot10^{-4}$ & $4\cdot10^{-6}$ & $4\cdot10^{-6}$  \\
\cline{3-7}
& & $P_{\textsc{MD}}$ & $3.45\cdot10^{-3}$ & $7.07\cdot10^{-3}$ & $6.47\cdot10^{-3}$ & $5.73\cdot10^{-3}$  \\
\cline{2-7}
& \multirow{2}*{11NN} & $P_{\textsc{FA}}$ & $3.39\cdot10^{-3}$ & $6.36\cdot10^{-3}$ & 0.055 & 0.03  \\
\cline{3-7}
& & $P_{\textsc{FA}}$ & 0.093 & 0.159 & 0.085 & 0.099 \\
\hline
\multirow{4}*{$\alpha^{\textsc{(II)}} = 0.8$} & \multirow{2}*{SVM} &  $P_{\textsc{MD}}$ & 0.994 & 0.991 & 0.992 & 0.992  \\
\cline{3-7}
& & $P_{\textsc{MD}}$ & $3.45\cdot10^{-3}$ & $7.07\cdot10^{-4}$ & $6.47\cdot10^{-3}$ & $5.73\cdot10^{-3}$  \\
\cline{2-7}
& \multirow{2}*{11NN} & $P_{\textsc{FA}}$ & 0.880 & 0.890 & 0.904 & 0.907  \\
\cline{3-7}
& & $P_{\textsc{MD}}$ & 0.130 & 0.128 & 0.114 & 0.101 \\
\hline
\end{tabular}
\end{center}
\end{table}

\begin{figure}[ht]
\centering
\includegraphics[width=0.45\textwidth]{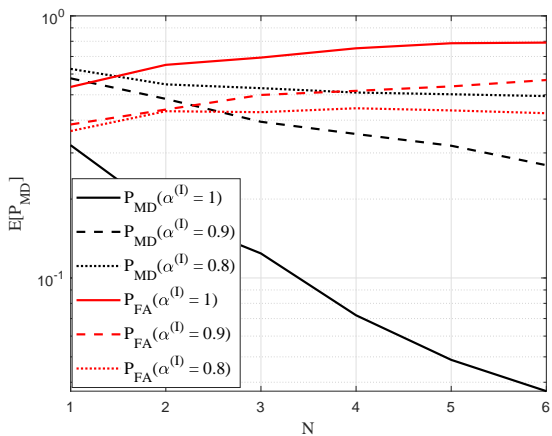}
\caption{Probabilities of FA and MD as function of the number of sub-carriers $N$, 11NN, with $\alpha^{\textsc{(II)}} = 0.9$, $\rho_{\textsc{AE}} = 0.8$, $\textsc{SNR}^{(\textsc{I})} = 15$dB, $\textsc{SNR}^{(\textsc{II})} = 20$dB and $M=100$. \label{fig:role_of_fad}}
\end{figure}

\begin{figure}[ht]
\centering
\includegraphics[width=0.45\textwidth]{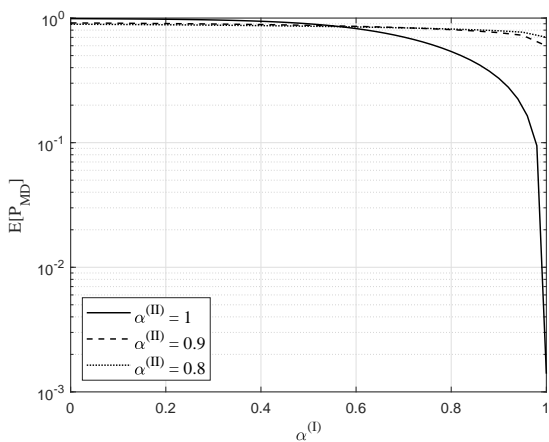}
\caption{Probability of MD as function of $\alpha^{\textsc{(I)}}$, LLR test, with different values of $\alpha^{\textsc{(II)}}$, $\rho_{\textsc{AE}} = 0.1$, $P_{\textsc{FA}} = 10^{-1}$, $\textsc{SNR}^{(\textsc{I})} = 15$dB, $\textsc{SNR}^{(\textsc{II})} = 20$dB. \label{fig:role_of_fad_stat}}
\end{figure}

Let us now assess the impact of the fading coefficient $\alpha$ in Phase 1 on the authentication performances. 
In Fig. \ref{fig:role_of_fad} we exploit a 11NN algorithm and we show that best performances are always obtained in absence of time-varying fading ($\alpha^{\textsc{(I)}} = 1$) during the training phase, outperforming even the case where the value of the time coefficient $\alpha^{\textsc{(I)}}$ remains the same during both phases.
Nevertheless this affects the false alarm probability, which experiences a slight worsening. In Fig. \ref{fig:role_of_fad_stat} we observe the same behavior described above for the LLR test, where we always obtain the lowest missed detection when considering $\alpha^{\textsc{(I)}} = 1$, regardless of value of $\alpha^{\textsc{(II)}}$.  

\subsection{Comparison of statistical and machine learning methods}

\begin{figure}[ht]
\centering
\includegraphics[width=0.45\textwidth]{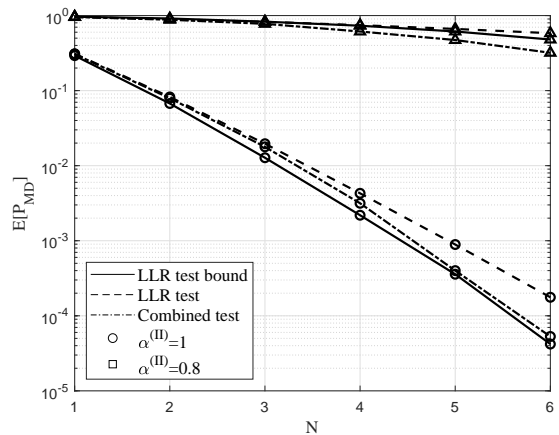}
\caption{Average MD probability $\mathbb{E}[P_{\textsc{MD}}]$ versus number of sub-carriers $N$, comparing statistical-based test methods, for different values of $\alpha^{\textsc{(II)}}$, $\alpha^{\textsc{(I)}} = 1$, $\rho_{\textsc{AE}} = 0.1$, with $\textsc{SNR}^{(\textsc{I})} = 15$dB and $\textsc{SNR}^{(\textsc{II})}= 20$dB.  \label{fig:confronto_stat}}
\end{figure}

In order to compare statistical and machine learning-based methods, let us assess the performance achieved by the statistical decision criteria presented in Section \ref{sec:stat_methods} in terms of average probability of MD. 
In Fig. \ref{fig:confronto_stat} the probability of FA has been fixed equal to $10^{\textsc{-4}}$ for both the examined methods, in absence of fading in the training phase ($\alpha^{\textsc{(I)}} = 1$) and with the spatial correlation coefficient $\rho_{\textsc{AE}}$ set to 0.1, with the meaning of Eve being very far from Bob's position. As a benchmark, we also show the curves obtained for the limit case presented in Sec \ref{subsec:bound}, which represents a lower bound on the performance achievable.  
From the results it is evident how much the channel variability (represented by decreasing values of $\alpha$) degrades the performance of the system, with a significant increase in the probability of MD with respect to the flat fading case (i.e., $\alpha^{\textsc{(II)}}=1$).
Looking at the figure, we note that, with respect to the single \ac{LLR} test, the combined test helps Bob to enhance the performance of the scheme, and this becomes more evident for increasing numbers of sub-carriers. 
In addition, we observe that with $\alpha^{\textsc{(II)}}=0.8$ the performance of the LLR test is very close to the bound, but both of them are outperformed by the combined test.

\begin{figure}
    \centering
    \includegraphics[width=0.45\textwidth]{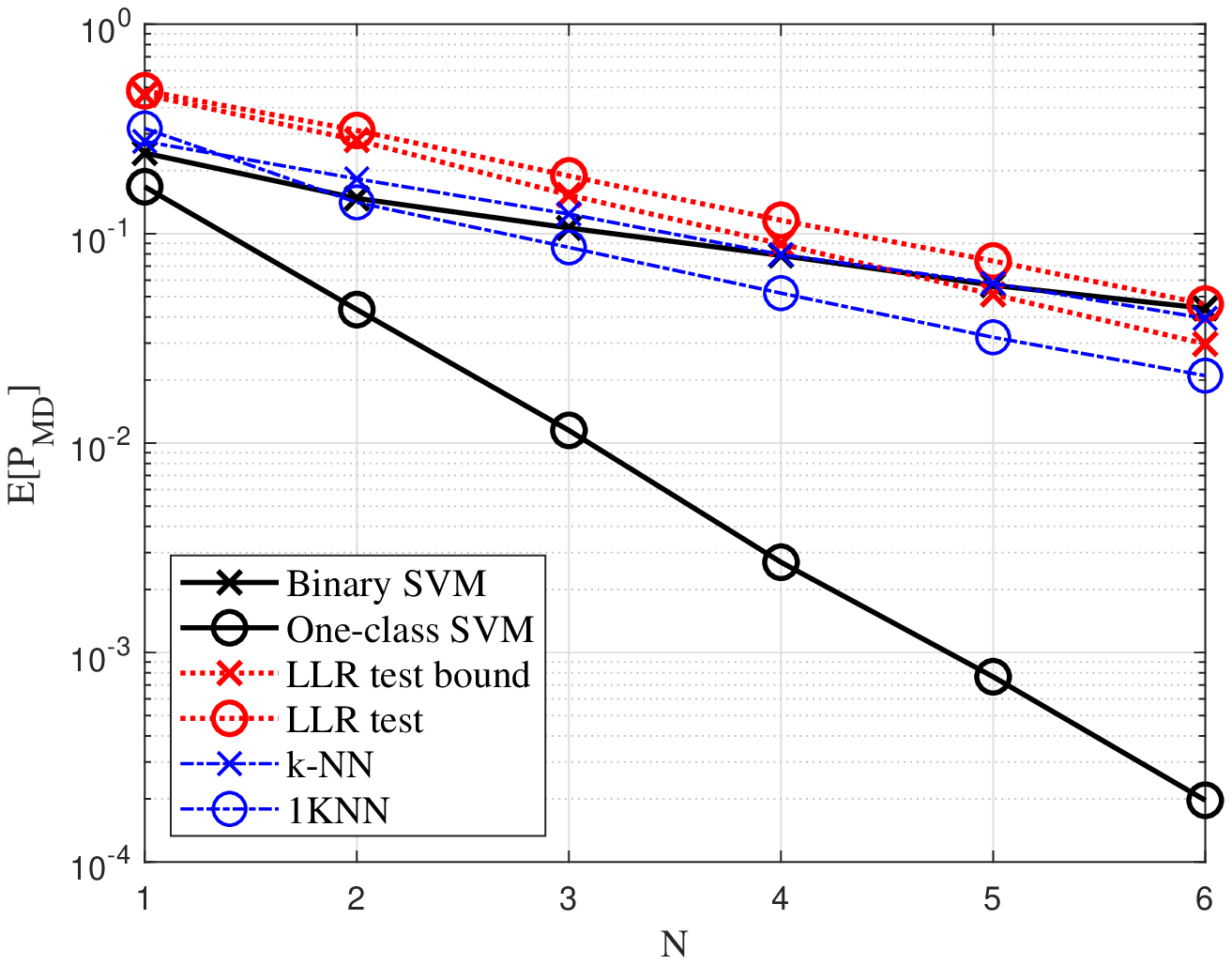}
    \caption{Average probability of MD as function of the number of sub-carriers $N$, with $\alpha^{\textsc{(I)}} = \alpha^{\textsc{(II)}} = 1$, $\rho_{\textsc{AE}} = 0.8$, $\textsc{SNR}^{(\textsc{I})} = 15$dB, $\textsc{SNR}^{(\textsc{II})} = 20$dB and $M=100$, considering $P_{\textsc{FA}}(\textsc{SVM}) = \left[ 
    1.39\cdot 10^{\textsc{-3}},  2.12\cdot 10^{\textsc{-4}},  6.85\cdot 10^{\textsc{-5}},  1.68 \cdot 10^{\textsc{-5}}, 5.55\cdot 10^{\textsc{-6}},  1.62\cdot 10^{\textsc{-6}} 
    \right]$, $P_{\textsc{FA}}(\textsc{1KNN}) = \left[ 
    6.76\cdot 10^{\textsc{-4}},  4.21\cdot 10^{\textsc{-4}},  <10^{\textsc{-6}},  <10^{\textsc{-6}}, <10^{\textsc{-6}},  <10^{\textsc{-6}} 
    \right]$ and $P_{\textsc{FA}}(\textsc{LLR}) = 10^{-4}$. }
    \label{fig:with_wo_eve}
\end{figure}

A similar assessment is reported in Fig. \ref{fig:with_wo_eve}, considering binary and one-class versions of \ac{SVM} and \ac{NN} algorithms besides LLR test and its bound. The \ac{LLR} test is equivalent to a one-class statistical method, while the bound corresponds to its binary version, since it considers the presence of Eve's samples in the training set.
As it results from the figure, in case of a high spatial correlation machine learning methods exhibit an opposite behavior with respect to the \ac{LLR} test with or without availability of Eve's samples: given almost the same $P_{\textsc{FA}}$, in fact, \ac{OCC} based on \ac{SVM} and 1KNN achieves a lower probability of MD with respect to their binary counterparts and the \ac{LLR} test.
This is probably due to the presence of negative samples highly correlated with the positive ones in the training set of the binary version, which ``confuses'' the algorithm and leads it to misclassify new instances with a high occurrence.

\begin{figure}[!t]
\begin{centering}
   \includegraphics[width=0.45\textwidth]{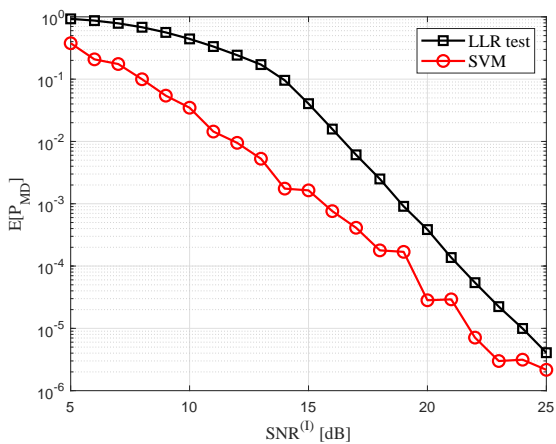}
\caption{
Average probability of MD using LLR test and SVM algorithm, with $N = 3$, $\rho_{\textsc{AE}}=0.5$, $M=100$, $\textsc{SNR}^{(\textsc{II})} = 20$dB, $\alpha^{\textsc{(I)}} = \alpha^{\textsc{(II)}} = 1$. $P_{\textsc{FA}} = [<10^{\textsc{-6}}, <10^{\textsc{-6}}, <10^{\textsc{-6}}, <10^{\textsc{-6}}, <10^{\textsc{-6}}, <10^{\textsc{-6}}, <10^{\textsc{-6}}, <10^{\textsc{-6}}, <10^{\textsc{-6}},$ $4.92\cdot10^{-6}, 7.73\cdot10^{-5}, 7.11\cdot10^{-4}, 4.48\cdot10^{-3}, 0.015, 0.0515, 0.117, 0.227, 0.3667, 0.512, 0.637, 0.761 ]$.}
\label{fig:PMD_vs_SNR_rev}
\par\end{centering}
\end{figure}

Further comparisons between statistical and machine learning based methods are discussed in the following.
In Fig. \ref{fig:PMD_vs_SNR_rev} we show how the \ac{SNR} in Phase I affects the authentication performance, exploiting both a LLR test and a SVM classifier. 
For a fair comparison, we fix as FA target for the LLR test the value of $P_{\textsc{FA}}$ achieved by the SVM classifier. It is possible to observe that the occurrence of MDs increases with the decreasing of the SNR (and the increasing of the variance); this can be explained by considering that for a low \ac{SNR} Bob is forced to accept a larger set of data in order to guarantee the desired level of FA. In any case, SVM outperforms the LLR test for all the considered values of $\textsc{SNR}^{(\textsc{I})}$.

\begin{figure}[!t]
\begin{centering}
   \includegraphics[width=0.45\textwidth]{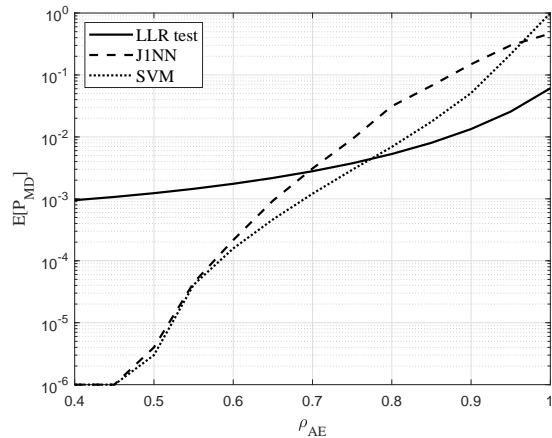}
\caption{Average probability of MD using LLR test, J1NN and SVM algorithm, with $N = 3$, $M=100$, $\textsc{SNR}^{(\textsc{I})} = 15$dB, $\textsc{SNR}^{(\textsc{II})} = 20$dB, $\alpha^{\textsc{(I)}} = 1$,$\alpha^{\textsc{(II)}} = 0.9$. $P_{\textsc{FA}} = 0.94$.}
\label{fig:PMD_vs_rho_rev}
\par\end{centering}
\end{figure}

In Fig. \ref{fig:PMD_vs_rho_rev} we assess performance achieved by different methods varying the attacker's spatial correlation $\rho_{\textsc{AE}}$, with time-varying fading $\alpha^{\textsc{(II)}} = 0.9$. It is evident, and will be confirmed in the following by looking at Tabs. \ref{tab:confronto_rho01} and \ref{tab:confronto_rho08}, that the use of machine learning approaches is convenient for small values of $\rho_{\textsc{AE}}$, while for larger values it is advisable to exploit statistical techniques.  

In Tables \ref{tab:confronto_rho01} and \ref{tab:confronto_rho08}\footnote{Probability less than $10^{-6}$ means that no error has been found over the entire data set. In order to perform an analytical evaluation, these values have been considered equal to $10^{-6}$.
} we show the performance achieved by \ac{OCC}, considering both \ac{NN} and \ac{SVM} algorithms, and we compare them with statistical methods applied to the same data set (for the sake of brevity we only report the results obtained by one among the four \ac{NN} classifiers) for increasing number of subcarriers (and thus for increasing number of features), with two different values of the spatial correlation parameter $\rho_{\textsc{AE}}$. 
In order to perform a fair comparison between different decision techniques, we choose as a target the probabilities of FA achieved by both the considered \ac{OCC} algorithms separately, and we compare the resulting probabilities of MD. The lowest probabilities of MD for each case are highlighted in the tables.
We observe that in general with low values of $\rho_{\textsc{AE}}$ \ac{NN} algorithms outperform \ac{SVM} and statistical methods, in terms of both $P_{\textsc{FA}}$ and $P_{\textsc{MD}}$. Excellent results are achieved especially in conditions of flat fading and with a large number of channels. 
When Eve is closer to Bob however, i.e. when the value of $\rho_{\textsc{AE}}$ is large, and with $\alpha^{\textsc{(II)}}$ different from 1, statistical methods (and the \ac{LLR} test in particular) maintain some advantage over the machine learning techniques.
In Tab. \ref{tab:confronto_rho08}, where $\rho_{\textsc{AE}} = 0.8$, \ac{SVM} algorithms achieve lower probabilities of \ac{MD} with respect to \ac{NN}, although they exhibit a higher $P_{\textsc{FA}}$ and therefore a worse balance between the two probabilities, as already shown in Fig. \ref{fig:confronto_a} and through $g_{mean}$ in Fig. \ref{fig:confronto_b}. 
We also note that, when the value of the spatial correlation remains constant, we obtain the same probability of MD for both \ac{SVM} and \ac{NN} algorithms, probably because $\rho_{\textsc{AE}}$ has no impact on the training phase. 
The same assertion holds true as regards the probability of FA achieved by \ac{SVM} algorithms when $\alpha^{\textsc{(II)}}$ does not vary, but not for \ac{NN} methods.

\begin{table*}[ht]
\begin{center}
\caption{Average MD and FA probabilities obtained by different test methods, for different values of $\alpha^{\textsc{(II)}}$, with $\alpha^{\textsc{(I)}} = 1$, $\rho_{\textsc{AE}} = 0.1$, $\textsc{SNR}^{(\textsc{I})} = 15$dB and $\textsc{SNR}^{(\textsc{II})} = 20$dB, $M=1000$. \label{tab:confronto_rho01}}
\begin{tabular}{|c|c|c|c|c|c|c|c|}
\hline
& $N$ & 1 & 2 & 3 & 4 & 5 & 6 \\
\hline
\hline
\multirow{8}*{$\alpha^{\textsc{(II)}}=1$} & $P_{\textsc{FA}}$ (1KNN) & 6.76$\cdot10^{\textsc{-4}}$ & $4.21\cdot10^{\textsc{-5}}$ & $<10^{\textsc{-6}}$ & $<10^{\textsc{-6}}$ & $<10^{\textsc{-6}}$ & $<10^{\textsc{-6}}$ \\
\cline{2-8}
& \cellcolor{ashgrey}{$P_{\textsc{MD}}$ (1KNN)} & \cellcolor{ashgrey}{0.055} & \cellcolor{ashgrey}{$4.74\cdot10^{\textsc{-4}}$} & \cellcolor{ashgrey}{$<10^{\textsc{-6}}$} & \cellcolor{ashgrey}{$<10^{\textsc{-6}}$} & \cellcolor{ashgrey}{$<10^{\textsc{-6}}$} & \cellcolor{ashgrey}{$<10^{\textsc{-6}}$} \\
\cline{2-8}
& $P_{\textsc{MD}}$ (LLR) & 0.255 & 0.094 & 0.044 & 0.012 & 0.0029 & $6.9\cdot10^{\textsc{-4}}$ \\
\cline{2-8}
& $P_{\textsc{MD}}$ (comb) & 0.254 & 0.089 & 0.036 & 0.010 & 0.0019 & $5.5\cdot10^{\textsc{-4}}$  \\
\cline{2-8}
\noalign{\vskip\doublerulesep
         \vskip-\arrayrulewidth}
\cline{2-8}
& $P_{\textsc{FA}}$ (SVM) & $1.39\cdot 10^{\textsc{-3}}$ & $2.12\cdot 10^{\textsc{-4}}$ & $6.85\cdot 10^{\textsc{-4}}$ & $1.68\cdot 10^{-5}$ & $5.55 \cdot 10^{\textsc{-6}}$ & $1.62\cdot10^{\textsc{-6}}$  \\
\cline{2-8}
& \cellcolor{ashgrey}{$P_{\textsc{MD}}$ (SVM)} & \cellcolor{ashgrey}{0.059} & \cellcolor{ashgrey}{$2.46 \cdot 10^{-3}$} & \cellcolor{ashgrey}{$<10^{\textsc{-6}}$} & \cellcolor{ashgrey}{$<10^{\textsc{-6}}$} & \cellcolor{ashgrey}{$<10^{\textsc{-6}}$} & \cellcolor{ashgrey}{$<10^{\textsc{-6}}$}  \\
\cline{2-8}
& $P_{\textsc{MD}}$ (LLR) & 0.233 & 0.073 & 0.021 & $6.6\cdot10^{\textsc{-3}}$ & $2\cdot10^{\textsc{-3}}$ & $6.09\cdot10^{\textsc{-4}}$  \\
\cline{2-8}
& $P_{\textsc{MD}}$ (comb) & 0.233 & 0.072 & 0.045 & 0.009 & 0.001 &  $7.7\cdot10^{-4}$ \\
\hline
\hline
\multirow{8}*{$\alpha^{\textsc{(II)}}=0.8$} & $P_{\textsc{FA}}$ & 0.525 & 0.632 & 0.745 & 0.847 & 0.930 & 0.949 \\
\cline{2-8}
& \cellcolor{ashgrey}{$P_{\textsc{MD}}$ (1KNN)} & \cellcolor{ashgrey}{0.055} & \cellcolor{ashgrey}{$4.74\cdot10^{\textsc{-4}}$} & \cellcolor{ashgrey}{$<10^{\textsc{-6}}$} & \cellcolor{ashgrey}{$<10^{\textsc{-6}}$} & \cellcolor{ashgrey}{$<10^{\textsc{-6}}$} & \cellcolor{ashgrey}{$<10^{\textsc{-6}}$} \\
\cline{2-8}
& $P_{\textsc{MD}}$ (LLR) & 0.176 & 0.058 & 0.016 & 0.0037 & $5.78\cdot10^{\textsc{-4}}$ & $1.71\cdot10^{\textsc{-4}}$ \\
\cline{2-8}
& $P_{\textsc{MD}}$ (comb) & 0.228 & 0.011 & 0.0027 & $3.9\cdot10^{\textsc{-5}}$ & $7\cdot10^{\textsc{-6}}$ &  $<10^{\textsc{-6}}$ \\
\cline{2-8}
\noalign{\vskip\doublerulesep
         \vskip-\arrayrulewidth}
\cline{2-8}
& $P_{\textsc{FA}}$ (SVM) & 0.830 & 0.953 & 0.981 & 0.988 & 0.989 & 0.990  \\
\cline{2-8}
& $P_{\textsc{MD}}$ (SVM) & 0.059 & $2.46\cdot10^{\textsc{-3}}$ & \cellcolor{ashgrey}{$<10^{\textsc{-6}}$} & \cellcolor{ashgrey}{$<10^{\textsc{-6}}$} & \cellcolor{ashgrey}{$<10^{\textsc{-6}}$} & \cellcolor{ashgrey}{$<10^{\textsc{-6}}$}  \\
\cline{2-8}
& $P_{\textsc{MD}}$ (LLR) & 0.054 & $4.9\cdot10^{\textsc{-3}}$ & $6.75\cdot10^{\textsc{-4}}$ & $5.57\cdot10^{\textsc{-5}}$ & $2.05\cdot10^{\textsc{-5}}$ & $<10^{\textsc{-6}}$  \\
\cline{2-8}
& $P_{\textsc{MD}}$ (comb) & \cellcolor{ashgrey}{0.043} & \cellcolor{ashgrey}{$1.07\cdot10^{-3}$} & $5.7\cdot10^{-5}$ & $10^{-6}$ & $<10^{-6}$ & $<10^{-6}$  \\
\hline
\end{tabular}
\end{center}
\end{table*}
\begin{table*}[ht]
\begin{center}
\caption{Average MD and FA probabilities obtained by different test methods, for different values of $\alpha^{\textsc{(II)}}$, with $\alpha^{\textsc{(I)}} = 1$, $\rho_{\textsc{AE}} = 0.8$, $\textsc{SNR}^{(\textsc{I})} = 15$dB and $\textsc{SNR}^{(\textsc{II})} = 20$dB, $M = 1000$. \label{tab:confronto_rho08}}
\begin{tabular}{|c|c|c|c|c|c|c|c|}
\hline
& $N$ & 1 & 2 & 3 & 4 & 5 & 6 \\
\hline
\multirow{8}*{$\alpha^{\textsc{(II)}} = 1$} & $P_{\textsc{FA}}$ & $6.76\cdot10^{\textsc{-4}}$ & $4.21\cdot10^{\textsc{-5}}$ & $<10^{\textsc{-6}}$ & $<10^{\textsc{-6}}$ & $<10^{\textsc{-6}}$ & $<10^{\textsc{-6}}$ \\
\cline{2-8}
& \cellcolor{ashgrey}{$P_{\textsc{MD}}$ (1KNN)} & \cellcolor{ashgrey}{0.318} & \cellcolor{ashgrey}{0.141} & \cellcolor{ashgrey}{0.086} & \cellcolor{ashgrey}{0.052} & \cellcolor{ashgrey}{0.032} & \cellcolor{ashgrey}{0.021} \\
\cline{2-8}
& $P_{\textsc{MD}}$ (LLR) & 0.531 & 0.378 & 0.319 & 0.183 & 0.099 & 0.052 \\
\cline{2-8}
& $P_{\textsc{MD}}$ (comb) & 0.592 & 0.380 & 0.279 & 0.168 & 0.077 & 0.047  \\
\cline{2-8}
\noalign{\vskip\doublerulesep
         \vskip-\arrayrulewidth}
\cline{2-8}
& $P_{\textsc{FA}}$ (SVM) & $1.39\cdot 10^{-3}$ & $2.12\cdot 10^{-4}$ & $6.85 \cdot 10^{-5}$ & $1.68 \cdot 10^{-5}$ & $5.55 \cdot 10^{\textsc{-6}}$ & $1.62 \cdot 10^{\textsc{-6}}$  \\
\cline{2-8}
& \cellcolor{ashgrey}{$P_{\textsc{MD}}$ (SVM)} & \cellcolor{ashgrey}{0.167} & \cellcolor{ashgrey}{0.044} & \cellcolor{ashgrey}{0.012} & \cellcolor{ashgrey}{$2.66 \cdot 10^{-3}$} & \cellcolor{ashgrey}{$6.83 \cdot 10^{-4}$} & \cellcolor{ashgrey}{$1.68 \cdot 10^{\textsc{-4}}$}  \\
\cline{2-8}
& $P_{\textsc{MD}}$ (LLR) & 0.494 & 0.313 & 0.189 & 0.120 & 0.074 & 0.047  \\
\cline{2-8}
& $P_{\textsc{MD}}$ (comb) & 0.495 & 0.316 & 0.188 & 0.113 & 0.068 & 0.044  \\
\hline
\hline
\multirow{8}*{$\alpha^{\textsc{(II)}} = 0.8$} & $P_{\textsc{FA}}$ & 0.684 & 0.867 & 0.918 & 0.955 & 0.974 & 0.983 \\
\cline{2-8}
& $P_{\textsc{MD}}$ (1KNN) & 0.318 & 0.141 & 0.086 & 0.052 & 0.032 & 0.021 \\
\cline{2-8}
& \cellcolor{ashgrey}{$P_{\textsc{MD}}$ (LLR)} & \cellcolor{ashgrey}{0.196} & \cellcolor{ashgrey}{0.052} & \cellcolor{ashgrey}{0.021} & \cellcolor{ashgrey}{0.007} & \cellcolor{ashgrey}{0.003} & \cellcolor{ashgrey}{0.001} \\
\cline{2-8}
& $P_{\textsc{MD}}$ (comb) & 0.319 & 0.124 & 0.073 & 0.040 & 0.023 & 0.016 \\
\cline{2-8}
\noalign{\vskip\doublerulesep
         \vskip-\arrayrulewidth}
\cline{2-8}
& $P_{\textsc{FA}}$ (SVM) & 0.830 & 0.953 & 0.981 & 0.988 & 0.989 & 0.990  \\
\cline{2-8}
& $P_{\textsc{MD}}$ (SVM) & 0.167 & 0.044 & 0.012 & $2.66 \cdot 10^{-3}$ & \cellcolor{ashgrey}{$6.83 \cdot 10^{-4}$} & \cellcolor{ashgrey}{$1.68 \cdot 10^{\textsc{-4}}$}  \\
\cline{2-8}
& $P_{\textsc{MD}}$ (LLR) & \cellcolor{ashgrey}{0.102} & \cellcolor{ashgrey}{0.017} & \cellcolor{ashgrey}{0.004} & \cellcolor{ashgrey}{$1.8 \cdot 10^{-3}$} & $1 \cdot 10^{-3}$ & $6.84 \cdot 10^{\textsc{-4}}$  \\
\cline{2-8}
& $P_{\textsc{MD}}$ (comb) & 0.166 & 0.043 & 0.018 & $9.82\cdot10^{-3}$ & $9.61\cdot10^{-3}$ & $7.75\cdot10^{-3}$   \\
\hline
\end{tabular}
\end{center}
\end{table*}

\begin{figure}[!t]
\begin{centering}
 \subfigure[]
   {\includegraphics[width=0.45\textwidth]{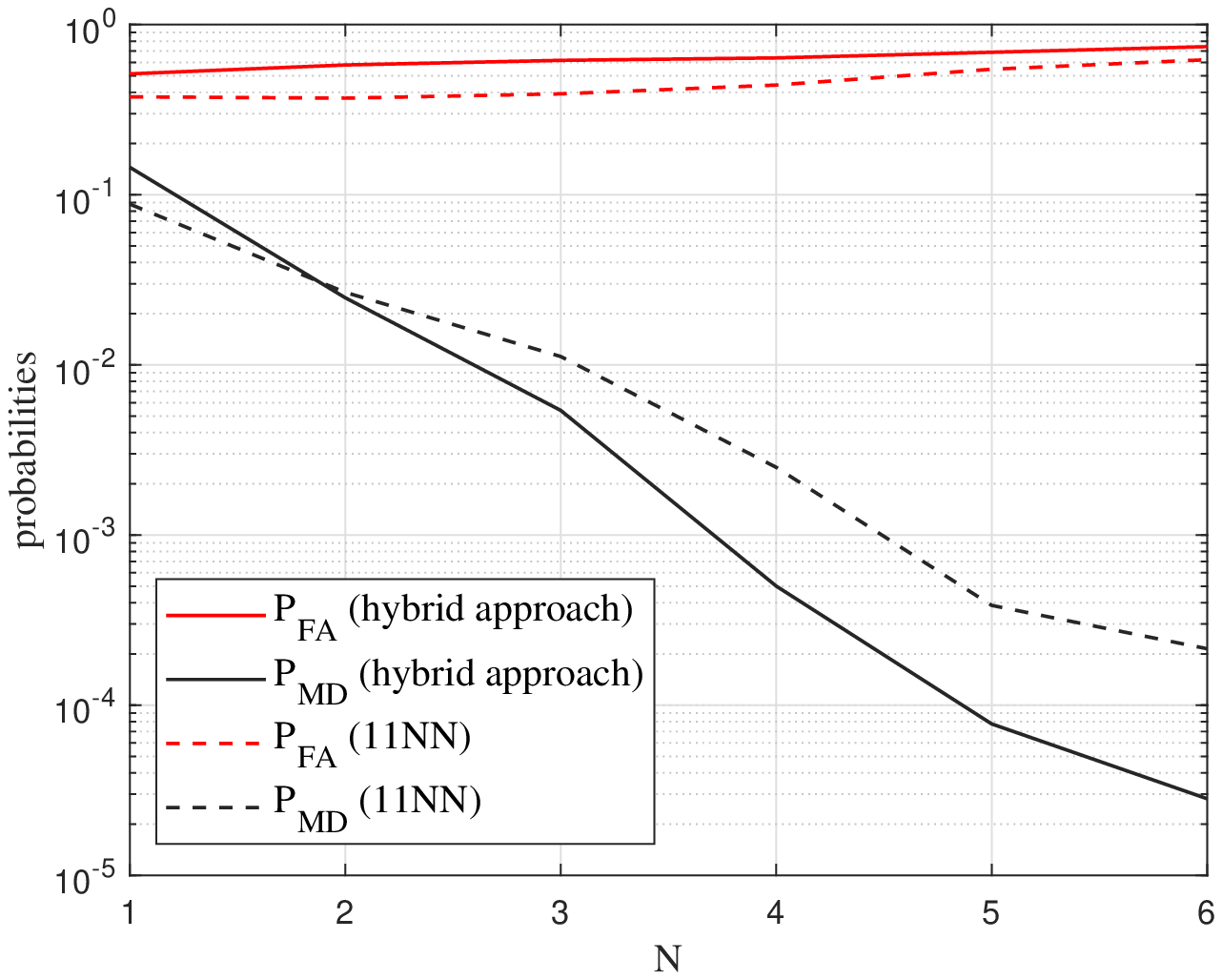}
   \label{fig:hybrid_a}}
 \subfigure[]
   {\includegraphics[width=0.45\textwidth]{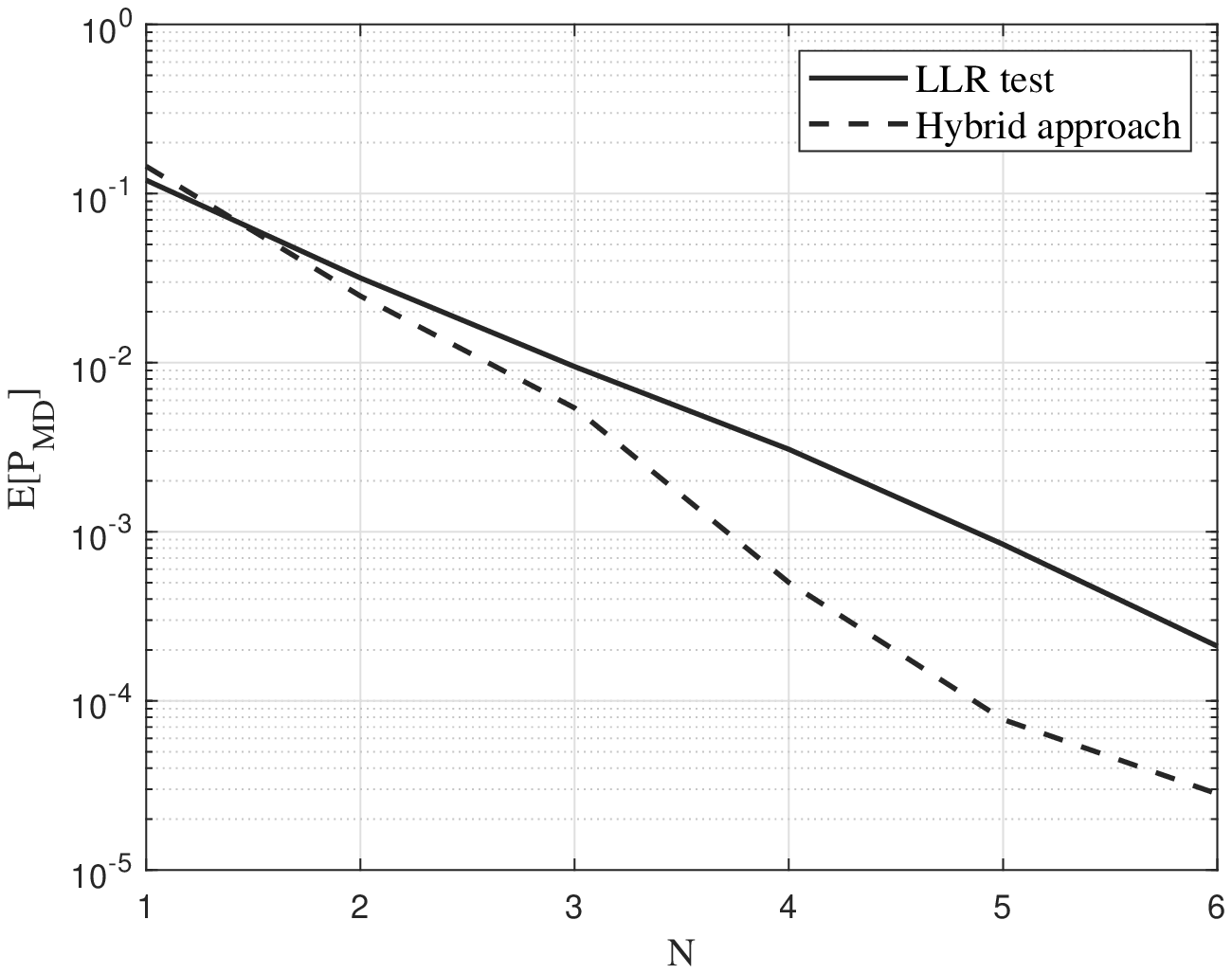}
   \label{fig:hybrid_b}}
\caption{
a) Probabilities of FA and MD obtained by using a hybrid approach and a 11NN with Euclidean metric, and b) Probability of MD achieved by LLR test and hybrid approach, with $M=100$, $\alpha^{\textsc{(I)}} = 1$, $\alpha^{\textsc{(II)}} = 0.9$, $\rho_{\textsc{AE}} = 0.1$, $\textsc{SNR}^{(\textsc{I})} = 15$dB, $\textsc{SNR}^{(\textsc{II})} = 20$dB. 
}
\label{fig:hybrid}
\par\end{centering}
\end{figure}

We finally test the proposed hybrid approach, comparing it with statistical and machine learning methods. In Fig. \ref{fig:hybrid_a} we report the results obtained by considering two different metrics in presence of time-varying fading $\alpha^{\textsc{(II)}}=0.9$. It is evident that using the LLR metric within a \ac{OCNN} algorithm (11NN in this particular case) improves the capability of the system to recognize forged messages coming from Eve, while it leads to a small loss in terms of false alarms raised, which becomes negligible with the increase of the number of subcarriers.
In Fig. \ref{fig:hybrid_b} we compare the probability of MD achieved by applying a statistical method which exploits a LLR test and the proposed hybrid approach, given the same probability of FA. Also in this case, the hybrid method has some advantage, especially with a large number of subcarriers.
We have therefore shown that the proposed approach improves authentication performance of both statistical techniques and machine learning methods based on NN algorithms. Nevertheless, statistical approaches maintain the advantage of having a very low complexity with respect to machine learning techniques.

\section{Conclusion}
\label{sec:concl}

We have assessed the performance achieved by different decision techniques in a physical layer authentication scenario with time-varying fading and in presence of an attacker. We have considered different methods based on both statistical criteria and machine learning algorithms.
We have shown that using a large training set that includes different realizations of the time-varying fading is not beneficial from the security point of view. We have also shown how clustering algorithms can help to avoid the use of higher layer authentication techniques in the initial phase, with only a small loss in terms of performance when a medium-large spatial correlation with the attacker channel exists. Somehow unexpectedly, in the same conditions the use of binary classification algorithms does not bring any advantage over their \ac{OCC} counterparts.
Our results demonstrate that \ac{NN} techniques are able to achieve a better trade-off between the \ac{FA} and \ac{MD} probabilities than the other considered classification methods. Moreover, they always result to be the best choice with low values of the spatial correlation, while in the other cases the application of statistical techniques leads to better performance.

\bibliographystyle{IEEEtran}
\bibliography{Archive}

\end{document}